\documentclass[a4paper,11pt]{article}
\pdfoutput=1
\usepackage{jcappub}
\usepackage[utf8x]{inputenc} 
\usepackage[normalem]{ulem}

\usepackage[dvipsnames]{xcolor}

\newcommand{\gbl}{g_\text{B3L}}
\newcommand{\mzbl}{m_\text{ZB3L}}

\newcommand{\lhs}{\lambda_\text{HS}}
\newcommand{\ls}{\lambda_S}
\newcommand{\lh}{\lambda_H}
\newcommand{\zbl}{Z_\text{B3L}}
\newcommand{\mpl}{M_\text{pl}}
\newcommand{\mdm}{m_\chi}
\usepackage{tcolorbox}
\usepackage[normalem]{ulem}

\begin{document}
\title{Shedding Flavor on Dark via Freeze-in: \\ $U(1)_{B-3L_i}$ Gauged Extensions}
\author[a]{Basabendu Barman,}
\author[b]{Purusottam Ghosh,}
\author[c]{\\Anish Ghoshal,}
\author[d]{Lopamudra Mukherjee}
\affiliation[a]{\,Centro de Investigaciones, Universidad Antonio Nari\~{n}o\\Carrera 3 este \# 47A-15, Bogot{\'a}, Colombia}
\affiliation[b]{\,Regional Centre for Accelerator-based Particle Physics, Harish-Chandra Research Institute,
HBNI, Chhatnag Road, Jhunsi, Allahabad - 211 019, India}
\affiliation[c]{\,Institute of Theoretical Physics, Faculty of Physics, University of Warsaw, \\ ul.  Pasteura 5, 02-093 Warsaw, Poland}
\affiliation[d]{\,Department of Physics and Astronomy, \\
225 Lewis Hall, University of Mississippi, Oxford, MS 38677-1848, USA}
\emailAdd{basabendu88barman@gmail.com}
\emailAdd{purusottamghosh@hri.res.in}
\emailAdd{anish.ghoshal@fuw.edu.pl}
\emailAdd{lmukherj@olemiss.edu}

\abstract{We consider a singlet fermionic dark matter (DM) $\chi$ in a gauged $U(1)_{B-3L_i}$ extension of the Standard Model (SM), with $i\in e\,,\mu\,,\tau$, and derive bounds on the allowed parameter space, considering its production via freeze-in mechanism. The DM communicates with the SM only through flavorful vector-portal $\zbl$ due to its non-trivial charge $x$ under $U(1)_{B-3L_{i}}$, which also guarantees the stability of the DM over the age of the Universe for $x\neq\{\pm 3/2,\pm 3\}$. Considering $\zbl$ to lie within the mass range of a few MeV up to a few GeV, we obtain constraints on the gauge coupling $\gbl$ from the requirement of producing right relic abundance. Taking limits from various (present and future) experimental facilities, e.g., NuCal, NA64, FASER, SHiP into account, we show that the relic density allowed parameter space for the frozen in DM can be probed with $\gbl\gtrsim 10^{-8}$ for both $\mdm<\mzbl/2$ and $\mdm\gtrsim\mzbl$, while $\gbl\lesssim 10^{-8}$ remains mostly unconstrained. We also briefly comment on the implications of neutrino mass generation via Type-I seesaw and anomalous $(g-2)_\mu$ in context with $B-3L_\mu$ gauged symmetry. 
}
\begin{flushright}
  PI/UAN-2022-709FT \\
  HRI-RECAPP-2021-013
\end{flushright}
\maketitle
\section{Introduction}
\label{sec:intro}

The minimal Standard Model (SM) of eletroweak and strong interactions has accidental global symmetries, namely the baryon number $B$ and the lepton number $L_i$, with $i\in e\,,\mu\,,\tau$. These symmetries are anomalous, and a common strategy for venturing to gauge theories beyond the SM is to gauge some linear combination of these symmetries which is anomaly-free. Almost four decades ago, it was observed that $U(1)_{B-L}$~\cite{Khalil:2008kp,Okada:2010wd,Kanemura:2011vm,Lindner:2011it,Okada:2012sg,Basak:2013cga,Kanemura:2014rpa,Okada:2016gsh,Biswas:2016ewm,Nanda:2017bmi,Singirala:2017cch,Bandyopadhyay:2017bgh,DeRomeri:2017oxa,Borah:2018smz,Okada:2018tgy,Das:2018tbd,Biswas:2019ygr,Das:2019pua,Baules:2019zwk,Mohapatra:2019ysk,Mohapatra:2020bze,Dutta:2020xwn,Biswas:2020ldp,Okada:2020evk,Ghosh:2021khk,Nath:2021uqb} symmetry is an anomaly free gauge extension of SM that treats all fermion generations on the same footing\footnote{Models where the generational universality is violated by gauging quantities like $L_{\mu}-L_{\tau}$ has received lots of interest in recent years~\cite{He:1991qd,PhysRevD.64.055006,Ma:2001md,Salvioni:2009jp,Heeck:2011wj,Harigaya:2013twa,CARONE2013118,Altmannshofer:2014cfa,Farzan:2015doa,Farzan:2015hkd,Biswas:2016yjr,Biswas:2019twf}.}. However, it is important to note that the new abelian gauge groups under which the SM fields carry non-trivial charges, are strongly constrained by the requirement of anomaly cancellation and the structure of the Cabibbo-Kobayashi-Maskawa (CKM) and Pontecorvo-Maki-Nakagawa-Sakata (PMNS) mixing matrices. These taken together into account, leave a limited number of global symmetries of the SM that can be gauged with the addition of only three right-handed neutrinos: $U(1)_{B-L}$, $U(1)_{B-3 L_{i}}$ and combinations of these groups. Observationally, as pointed out in~\cite{Bauer:2020itv}, the flavour changing coupling of the new gauge boson to active neutrinos in the mass basis can distinguish between $U(1)_{B-L}$ and $U(1)_{B-3 L_{i}}$, albeit quark and charged lepton flavour transitions are strongly suppressed irrespective of the gauge group. Gauged $U(1)_{B-3 L_{i}}$ symmetry (where $i\in e\,,\mu\,,\tau$), has also been discussed in the context of neutrino mass generation, low-energy unification and flavour anomalies~\cite{Ma:1997nq,Ma:1998dr, Chang:2000xy,Pal:2003ip,Chang:2009tx,Lee:2010hf,Okada:2012sp,Chun:2018ibr,Allanach:2018lvl,Borah:2021jzu,Borah:2021khc}. 

Moreover, the failure of the SM to provide a viable dark matter (DM) candidate, whose existence in the Universe has already been proven unequivocally from several astrophysical~\cite{Zwicky:1933gu, Zwicky:1937zza, Rubin:1970zza, Clowe:2006eq} and cosmological~\cite{Hu:2001bc, Aghanim:2018eyx} evidences (for a review, see, e.g. Refs.~\cite{Jungman:1995df, Bertone:2004pz, Feng:2010gw}), compels one to explore beyond the realms of the SM. The weakly interacting massive particle (WIMP) paradigm is the most elusive one in explaining the observed relic abundance of the DM in the Universe, as the observed DM abundance is set typically by the weak scale interaction strength. However, the absence of significant excess, in typically nuclear recoil experiments~\cite{XENON:2018voc, XENON:2020kmp, PandaX:2018wtu}, is constantly cornering the WIMP paradigm. Out of several alternate prescriptions, the observed DM abundance can be generated by what is known as the {\it freeze-in}~\cite{McDonald:2001vt, Hall:2009bx, Chu:2011be, Bernal:2017kxu, Duch:2017khv, Biswas:2018aib,  Heeba:2018wtf, Barman:2019lvm, Barman:2021lot,Barman:2022njh}. The DM particles, in this case, are produced by the decay or annihilation of the bath particles. The production ceases due to cooling of the bath temperature below the relevant mass scale connecting the DM to the visible sector. As a result, in contrast to the WIMP scenario, in freeze-in, the DM abundance increases with the increase in interaction strength between the DM with the visible sector. Even though freeze-in is perfectly capable of explaining the whole DM abundance, it is challenging to probe because of its feeble coupling with the visible sector. In context of $U(1)_{B-L}$, DM phenomenology has been studied considering one of the right handed neutrinos, required for anomaly cancellation, to be a potential DM candidate in, for example, Refs.~\cite{Burell:2011wh,Okada:2012sg,Okada:2016gsh,Okada:2020cue,Nath:2021uqb} including particularly that of an ad-hoc singlet fermion as DM~\cite{Mohapatra:2019ysk, delaVega:2021wpx, Ghoshal:2022zwu}. In all of these scenarios, the interaction between DM and the SM particles is primarily mediated by the $B-L$ gauge boson. Hence, these scenarios predict a one-to-one correspondence between the experimental searches for the new gauge boson and DM physics.

Motivated from these, in this work we have explored the possibility of probing freeze-in DM scenario in gauged $B-3L_i$ with a singlet vector-like fermionic DM charged under the new gauge symmetry. As the DM is vector-like, it does not contribute to the gauge anomaly that leaves us with the opportunity to consider its charge as a free parameter. The charge assignment allows the DM to talk to the SM {\it only} via the neutral gauge boson, also ensuring its stability over the lifetime of the Universe. As a result, the DM parameter space can be constrained from experimental bounds on the mass and coupling of the new gauge boson, establishing a direct connection between DM and gauge boson search. Also note the absence of an ad-hoc stabilizing $Z_2$ symmetry, which is very common in $B-L$ gauge extensions with one of the RHNs as a DM candidate (for example, in~\cite{Okada:2016gsh}). Moreover, considering one of the RHNs to be DM, the model albeit becomes minimal, but one has to take into account the effect of the Higgs portal, which we do not intend to, as we aim to investigate the status of freeze-in parameter space in the light of massive gauge boson searches. The requirement of gauge anomaly cancellation also introduces a heavy singlet right handed neutrino (RHN) which, along with one or two more right handed neutrinos having vanishing $B-3L_i$ charges, can take part in generating light neutrino masses through usual Type-I seesaw mechanism. We typically focus on the regime where the massive gauge boson of the local $B-3L_i$ symmetry is sub-GeV\footnote{Inflationary dynamics of such light mediators were studied in Ref.~\cite{Paul:2018njm} with constraints from $N_\text{eff}$ observables in~\cite{Paul:2021ewd}.} and can be as heavy as $\mzbl\lesssim 10$ GeV. We show, for suitable choices of the DM charge $x$ under $B-3L_i$ symmetry, together with the gauge coupling $\gbl$, it is possible to produce observed relic abundance for the DM via freeze-in mechanism in the early Universe. The central idea of this study is to establish the fact that although the size of the gauge coupling is small enough, preventing the DM to achieve thermal-equilibrium with the SM bath, such small coupling still remains within the reach of various current and future planned/proposed experiments that typically search for very weakly coupled light states.

The paper is organized as follows. The model set-up is discussed in Sec.~\ref{sec:model}, where we tabulate the particle content mentioning their charges, together with the relevant Lagrangian in subsection.~\ref{sec:content}; the light neutrino mass generation via Type-I seesaw is discussed in subsection.~\ref{sec:type-1}; we then move on to the calculation of DM abundance via freeze-in in Sec.~\ref{sec:dm-pheno}, where we elaborate the implementation of coupled Boltzmann equation in subsection~\ref{sec:fi-yld} following the discussion about the DM parameter space in subsection.~\ref{sec: paramspace}, mentioning all the relevant constraints. Finally, we conclude in Sec.~\ref{sec:concl}.

\section{Model Framework}
\label{sec:model}
\begin{table}[htb!]
\begin{center}
\begin{tabular}{|c|c|c|c|c|c|c|c|c|c|c|}
\hline
Fields &$SU(3)_c$ & $SU(2)_L$ & $U(1)_Y$ & $U\left(1\right)_{B-3L_i}$ \\ [0.5ex] 
\hline
\hline
$Q_L^i$ & 3 & 2 & 1/6 & 1/3 \\
$u_R^i$ & 3 & 1 & 2/3 & 1/3 \\
$d_R^i$ & 3 & 1 & -1/3 &  1/3 \\
\hline\hline
$L^e$ & 1 & 2 & -1/2 & -3\,$\delta_{ei}$ \\
$L^\mu$ & 1 & 2 & -1/2 & -3\,$\delta_{\mu i}$ \\
$L^\tau$ & 1 & 2 & -1/2 & -3\,$\delta_{\tau i}$ \\
\hline\hline
$e_R^e$ & 1 & 1 & -1 &  -3\,$\delta_{ei}$ \\
$e_R^\mu$ & 1 & 1 & -1 &  -3\,$\delta_{\mu i}$ \\
$e_R^\tau$ & 1 & 1 & -1 &  -3\,$\delta_{\tau i}$ \\
\hline\hline
$N_R^e$ & 1 & 1 & 0 & -3\,$\delta_{ei}$ \\
$N_R^\mu$ & 1 & 1 & 0 & -3\,$\delta_{\mu i}$ \\
$N_R^\tau$ & 1 & 1 & 0 & -3\,$\delta_{\tau i}$ \\
\hline\hline
$H$ & 1 & 2 & 1/2 & 0 \\
$S$ & 1 & 1 & 0 & 3 \\
\hline\hline
$\chi$ & 1 & 1 & 0 & $x$ \\
\hline
\end{tabular}
\end{center}
\caption {\it Charges of the new particles under $\text{SM}\times (B-3L_i)$ symmetry, where $i\in e\,,\mu\,,\tau$ and $\delta_{ij}=1$ for $i=j$, else zero.}
\label{tab:particles1}
\end{table}

We consider three beyond the SM (BSM) scenarios: the SM augmented via a new $U(1)_X$ gauge group, with $X\equiv B-3L_i\,;\,i\in e\,,\mu\,,\tau$. The generation dependent gauge charge assignment of the SM fields lead to triangle anomalies, that require extra fields to cancel them which carry charge under $U(1)_{B-3L_i}$. The condition for anomaly cancellations come from the triangle diagrams with gauge bosons at the vertices:
\begin{equation}
\sum_{\text{LH}}{\rm Tr}\left[T^a\{T^b,T^c\}\right]-\sum_{\text{RH}}{\rm Tr}\left[T^a\{T^b,T^c\}\right]\,,    
\end{equation}
with ``Tr" standing for trace and $T^{a,b,c}$ denoting the corresponding generators. This vanishes exactly for the vector-like states as they contribute identically to the left and right-handed representation. However, the contributions from other fermions do not vanish in general due to their chiral nature. The $B-3L_i$ charge assignment in Tab.~\ref{tab:particles1} satisfies cancellation of the gauge and the gravitational anomalies that can be trivially checked.



\subsection{Particle content}
\label{sec:content}
As mentioned above, in order to cancel the gauge anomalies we introduce three right handed neutrions (RHN), singlet under the SM gauge symmetry, but carrying non-trivial charges depending on the nature of $U(1)_{B-3L_i}$. These RHNs take part in the anomaly cancellation as mentioned before. We further introduce a vector-like singlet fermion, carrying some charge $x$ under the new gauge symmetry. This can serve as a potential DM candidate in our model. The charge $x$ assigned to the DM is arbitrary, however $x\neq\{\pm\frac{3}{2}\,,\pm 3\}$  such that the DM communicates to the SM {\it only} through the new gauge boson. The same charge assignments also render the DM stable by forbidding the Yukawa interactions, e.g., $\overline{\chi}\,e_R\,S\,,\overline{\chi}\,N_R\,S\,,\overline{L}\,H\,\chi$ etc. Thus, the DM in the present set-up can be made absolutely stable without introducing any ad-hoc discrete symmetry. The full particle spectrum, together with the charges under different gauge groups are summarized in Tab.~\ref{tab:particles1}. With this particle content at our disposal, we can write the following interaction Lagrangian
\begin{equation}
\mathcal{L}^{{B-3L_i}} \supset\mathcal{L}_\text{gauge} + \mathcal{L}_\text{scalar}+\mathcal{L}_\text{fermion}^{{B-3L_i}} +\mathcal{L}_\text{yuk}^{{B-3L_i}}\,, 
\label{eq:lgrng}
\end{equation}
with
\begin{equation}
\mathcal{L}_\text{gauge}\supset -\frac{1}{4}\,Z_{\text{B3L}_{\mu\nu}}\,Z^{\text{B3L}{\mu\nu}} - \frac{\epsilon}{2}\,Z_{\text{B3L}\mu\nu}\,B^{\mu\nu}\,,  
\label{eq:lg}
\end{equation}
where we consider $\epsilon=0$ at the scale of $B-3L_i$ breaking~\cite{Okada:2018ktp}. The Lagrangian corresponding to the scalar sector is given by
\begin{equation}
\mathcal{L}_\text{scalar}=\left|D_\mu\,S\right|^2-V(H,S)\,, 
\label{eq:ls}
\end{equation}
where the scalar potential reads
\begin{equation}
V(H,S) = -\mu_H^2\,\left|H\right|^2 + \lh\,\left|H\right|^4 - \mu_S^2\,\left|S\right|^2 + \ls\,\left|S\right|^4 + \lhs\,\left|H\right|^2\,\left|S\right|^2\,,
\label{eq:pot}
\end{equation}
and the covariant derivative reads
\begin{equation}
D_\mu\equiv \partial_\mu-i\,g_2\,\tau^a\,W_\mu^a-i\,g_1\,Y\,B_\mu-i\,\gbl\,Y_\text{B3L}\,Z_{\text{B3L}\mu}\,,  
\end{equation}
where $g_{1,2}$ corresponds to the $U(1)_Y$ and $SU(2)_L$ gauge couplings respectively and $\gbl$ is the new gauge coupling and $Y_\text{B3L}$ is the corresponding charge as mentioned in Tab.~\ref{tab:particles1}. The $SU(2)_L$ doublet $H$ mixes with the singlet $S$ once the fields acquire non-zero vacuum expectation values (VEV). The two scalars in the unitary gauge can be parametrized as
\begin{equation}
H=
\begin{pmatrix}
0 \\ \frac{1}{\sqrt{2}}(h+v_d)
\end{pmatrix}\,;~S=\frac{1}{\sqrt{2}}\,\left(s+v_s\right)\,,
\end{equation}
with $v_{d,s}$ being the corresponding VEVs. On minimizing $V(H,S)$ we obtain
\begin{align}
&\mu_H^2 = v_d^2\,\lambda_H+\frac{1}{2}\,\lambda_{HS}\,v_s^2\,
\nonumber\\&
\mu_S^2 = v_s^22\,\lambda_S + \frac{1}{2}\,\lambda_{HS}\,v_d^2\,.    
\end{align}
The CP even mass matrix reads
\begin{equation}
\mathcal{M}_\text{CPE}^2=\left(
\begin{array}{cc}
 2 v_d^2\,\lambda_H & v_d\,v_s\,\lambda_{HS}\\
 v_d\,v_s \,\lambda_{HS} & 2\,v_s^2\,\lambda_S\\
\end{array}
\right)\,,
\end{equation}
with eigenvalues
\begin{equation}
m_{1,2} = \frac{1}{2}\,\Bigl(A+C\mp\sqrt{A^2+4\,B^2-2\,A\,C+C^2}\Bigr)\,,    
\end{equation}
with $A=2\,v_d^2\,\lambda_H\,,B=v_d\,v_s\,\lambda_{HS}$ and $C=2\,v_s^2\,\lambda_S$. We can diagonalize this mass matrix using a $2\times2$ rotation matrix that leaves two physical eigenstates in the mass basis
\begin{equation}
\begin{pmatrix}
h \\ s
\end{pmatrix}=
\begin{pmatrix}
c_\theta & -s_\theta \\ s_\theta & c_\theta
\end{pmatrix}\,
\begin{pmatrix}
h_1 \\ h_2
\end{pmatrix}\,,
\end{equation}
where $h_1$ is SM-like Higgs with mass $\sim 125$ GeV and the mixing angle is given by
\begin{equation}
\sin2\theta = \frac{2\,v_d\,v_s}{m_2^2-m_1^2}\,\lambda_{HS}\,. \label{eq:mix}   
\end{equation}
On spontaneous breaking of $U(1)_{B-3L_i}$, we get a massive neutral gauge boson with mass
\begin{equation}
\mzbl = 3\,\gbl\,v_s\,.  
\end{equation}
One may then trade-off the renormalizable couplings appearing in $V(H,S)$ with the physical masses and mixings
\begin{align}\label{eq:cplngs}
&\ls=\frac{1}{2\,v_s^2}\,\left(m_1^2\,\sin ^2\theta+m_2^2\,\cos ^2\theta\right)
\nonumber\\&
\lh=\frac{1}{2\,v_d^2}\,\left(m_1^2\,\cos ^2\theta+m_2^2\,\sin^2\theta\right)
\nonumber\\&
\lhs=\frac{\sin2\theta}{2\,v_d\,v_s}\, \left(m_1^2-m_2^2\right)\,.
\end{align}
In presence of the mixing between the two scalars, the tree level interactions of the SM Higgs with the SM fermion and gauge bosons get modified. In such
a scenario, the LHC sets a bound on the mixing $|\sin\theta|\leq 0.36$, from the measurement of signal strength in the di-photon channel~\cite{Robens:2015gla,Robens:2016xkb,Chalons:2016jeu}. The strongest bound on singlet scalar-SM Higgs mixing appears from $W$-boson mass correction at NLO~\cite{Lopez-Val:2014jva} for the mass range $250\lesssim m_2\lesssim 850$ GeV for $0.2\lesssim\sin\theta\lesssim 0.3$. For $m_2>$ 850 GeV, the bounds from the requirement of perturbativity
and unitarity turn out to be dominant, requiring $\sin\theta\lesssim 0.2$. For $m_2 <$ 250 GeV, the LHC and LEP direct search and measured Higgs signal strength restrict the mixing angle $\sin\theta\lesssim 0.25$~\cite{CMS:2015hra,Strassler:2006ri}.  Also, the SM Higgs in the present model can decay into a pair of $\zbl$, which is tightly constrained from the measurement of Higgs invisible decay branching that demands $\text{Br}(h\to\text{invisible})<0.11$~\cite{CMS:2016dhk}. This bound, however, is trivially satisfied since $\Gamma(h_1\to\zbl\zbl
)\propto \sin^2\theta \gbl^2$ and $\gbl\sim\mathcal{O}(10^{-8})$, required for freeze-in as we shall see in Sec.~\ref{sec:dm-pheno}. In the entire analysis we shall fix $\sin\theta=10^{-2}$ and the mass of the non-standard Higgs to $m_2=500$ GeV. Note that, due to large $v_s$, both the scalar couplings $\lhs$ and $\ls$ are of the same order as that of $\gbl$. 

The Lagrangian for the fermion sector under the $U(1)_{B-3L_i}$ turns out to be
\begin{equation}
\mathcal{L}_\text{fermion}^{{B-3L_i}} \supset\overline{N_R}\,i\gamma^\mu\,D_\mu\,N_R+\overline{\chi}\,i\gamma^\mu\,D_\mu\,\chi- \sum_{i\neq j,k}\frac{1}{2}\,(M_N)_{jk}\,\overline{N_R^c}^j\,N_R^k-\mdm\,\overline{\chi}\,\chi ,
\label{eq:lf}
\end{equation}
where we can write a bare mass term $m_\chi$ for the DM, without violating the gauge symmetry, thanks to its vector-like transformation.

Finally, the Yukawa interactions are given by
\begin{equation}
-\mathcal{L}_\text{yuk}^{{B-3L_i}} \supset \sum_{j \neq i} \frac{1}{2}\,y_{ij}\,\overline{N_R^c}^i\,N_R^j\,S+ \sum_{\alpha \neq i}\,\sum_{j\neq i}\,y_{\alpha j}\,\overline{L_\alpha}\,\widetilde{H}\,N_{R_j} +  y_{L_i}\,\overline{L^i}\,\widetilde{H}\,N_R^i+ \text{H.c}\,,
\label{eq:lyuk}
\end{equation}
where $\alpha\in e\,,\mu\,,\tau$, $j\in e\,,\mu\,,\tau$ . Before moving on, it is important to note that the phenomenology of the present set-up is governed by the following set of independent parameters $\{x\,,\gbl\,,\mzbl\,,\mdm\}\,$. As we shall see in Sec.~\ref{sec:dm-pheno}, the choice of the charge $x$ and the gauge coupling $\gbl$ crucially determine the detection prospects for freeze-in.  

\subsection{Generation of the SM neutrino mass}
\label{sec:type-1}
In the present set-up the light neutrino mass $m_\nu$ can be generated via standard Type-I seesaw~\cite{ PhysRevLett.44.912, Gell-Mann:1979vob} mechanism as
\begin{equation}
m_\nu=-M_D\,(M_R)^{-1}\,M_D^T\,,
\end{equation}
where $M_D$ and $M_R$ can be derived from  the Yukawa interaction of SM leptons in Eq.~\eqref{eq:lf} and singlet fermions in Eq.~\eqref{eq:lyuk}. Depending on the charge of the $B-3L_i$ gauge symmetry, these matrices can be expressed as follows
\begin{align}
M_D^e = 
\begin{pmatrix}
y_{ee}\,\frac{v_d}{\sqrt{2}} & 0 & 0 \\ 0 & y_{\mu\mu}\,\frac{v_d}{\sqrt{2}} & y_{\mu3}\,\frac{v_d}{\sqrt{2}} \\0 & y_{\tau2}\,\frac{v_d}{\sqrt{2}} & y_{\tau\tau}\,\frac{v_d}{\sqrt{2}}\ 
\end{pmatrix}\,;
M_R^e=
\begin{pmatrix}
0 & y_{12}\,\frac{v_s}{\sqrt{2}} & y_{13}\,\frac{v_s}{\sqrt{2}} \\
y_{21}\,\frac{v_s}{\sqrt{2}} & M_{22} & M_{23}\\
y_{31}\,\frac{v_s}{\sqrt{2}} & M_{32} & M_{33}
\end{pmatrix}\,,
\label{eq:neu-mat1}
\end{align}

\begin{align}
M_D^\mu = 
\begin{pmatrix}
y_{ee}\,\frac{v_d}{\sqrt{2}} & 0 & y_{e\tau}\,\frac{v_d}{\sqrt{2}} \\ 0 & y_{\mu\mu}\,\frac{v_d}{\sqrt{2}} & 0 \\ y_{\tau e}\,\frac{v_d}{\sqrt{2}} & 0 & y_{\tau\tau}\,\frac{v_d}{\sqrt{2}}\ 
\end{pmatrix}\,;
M_R^\mu =
\begin{pmatrix}
M_{11} & y_{12}\,\frac{v_s}{\sqrt{2}} & M_{13} \\
y_{21}\,\frac{v_s}{\sqrt{2}} & 0 & y_{23}\,\frac{v_s}{\sqrt{2}}\\
M_{31} & y_{32}\,\frac{v_s}{\sqrt{2}} & M_{33}
\end{pmatrix}\,,
\label{eq:neu-mat2}
\end{align}

\begin{align}
M_D^\tau = 
\begin{pmatrix}
y_{ee}\,\frac{v_d}{\sqrt{2}} & y_{e\mu}\,\frac{v_d}{\sqrt{2}} & 0 \\ y_{\mu e}\,\frac{v_d}{\sqrt{2}} & y_{\mu\mu}\,\frac{v_d}{\sqrt{2}} & 0 \\ 0 & 0 & y_{\tau\tau}\,\frac{v_d}{\sqrt{2}}\ 
\end{pmatrix}\,;
M_R^\tau =
\begin{pmatrix}
M_{11} & M_{12} & y_{13}\,\frac{v_s}{\sqrt{2}} \\
M_{21} & M_{22} & y_{23}\,\frac{v_s}{\sqrt{2}}\\
y_{13}\,\frac{v_s}{\sqrt{2}} & y_{23}\,\frac{v_s}{\sqrt{2}} & 0
\end{pmatrix}\,.
\label{eq:neu-mat3}
\end{align}
The light neutrino mass matrix is diagonalized utilizing the PMNS matrix $U_\text{PMNS}$ as
\begin{equation}
m_\nu = U_\text{PMNS}\,.m_\nu^\text{diag}\,.U_\text{PMNS}^T\,,    \end{equation}
where $m_\nu^\text{diag}\equiv\text{diag}\left(m_1\,,m_2\,,m_3\right)$ and the PMNS matrix is parametrized as
\begin{equation}
 U_\text{PMNS}=\left(
\begin{array}{ccc}
 c_{12}\,c_{13} & s_{12}\,c_{13} & s_{13} \\
 -s_{12}\,c_{23}-c_{12}\,s_{23}\,s_{13} & c_{12}\,c_{23}-s_{12} \,s_{23}\,s_{13} & s_{23}\,c_{13} \\
 s_{12}\,s_{23}-c_{12}\,c_{23}\,s_{13} & -s_{12}\,c_{23} s_{13}-c_{12}\,s_{23} & c_{23}\,c_{13} \\
\end{array}
\right)\,,    
\end{equation}
where $c_{ij}\equiv\cos\theta_{ij}$ is the cosine of the mixing angles. We also assume the charge lepton mass matrix to be diagonal. A detailed study of neutrino mass and mixing, appropriately fitting the observed data, has been discussed in~\cite{Wang:2019byi} in the context of $U(1)_{B-3L_i}$. Here we do not go into such details, we only perform a simple estimation to obtain the possible mass range of the RHNs that can provide the light neutrino mass in the correct ballpark. The estimation of possible RHN mass range is necessary in order to calculate the DM relic abundance in the next section. Neutrino oscillation experiments are sensitive to the mass-squared differences and the mixing angles, and the value of these parameters used in the analysis in the $3\sigma$ range are as follows~\cite{Zyla:2020zbs}
\begin{align}
&|\Delta m^2_\text{sol}| \equiv \left|m^2_2 -m^2_1 \right|
	\in [6.79-8.01]\times 10^{-5}  {\text{eV}^2} \,, 	\left|\Delta m^2_\text{atm}\right|\equiv |m^2_3 -m^2_1|
	\in [2.35-2.54]\times 10^{-3} {\text{eV}^2}\,,
	\nonumber\\& 
	~~~~~~~~~~~~~~~~\sin^2\theta_{12}\in [0.27,0.35] \,,
	\sin^2\theta_{23} \in [0.43,0.60] \,,
	\sin^2\theta_{13}\in [0.019,0.024]\,.
	\label{obs_para}
\end{align}
Since the sign of $\Delta m_{31}^2$ is undetermined, distinct neutrino mass hierarchies are possible. This, in turn, puts bound on model parameters, particularly, on the VEV $v_s$ and relevant Yukawa couplings. We consider the lightest active neutrino to be massless such that $\Delta m_\text{sol}^2\equiv m_2^2$ and $\Delta m_\text{atm}^2\equiv m_3^2$. We choose the case of $U(1)_{B-3L_e}$ for illustration, while for the other two gauge groups similar argument follows. 

We choose $y_{\mu\mu}=y_{\mu3}=y_{\tau2}=y_{\tau\tau}\equiv y_\ell$, $y_{12}=y_{13}\equiv y$ and $M_{22}=M_{33}\equiv M$. Now, for $v_s=\mzbl/\gbl=10^{10}$ GeV, light neutrino mass can be generated for  $y_\ell\sim 10^{-7}$, $y_{ee}\sim 10^{-2}$, $y\sim 10^{-2}$ with $M=1$ TeV. Even if we consider $M=10$ TeV, we can still obtain the light neutrino mass, by properly tuning the couplings.  However, in that case, the RHNs are beyond the present collider reach. Since the Yukawas do not determine the outcome of the DM phenomenology, we can safely fix them to some benchmark values such that light neutrino mass is satisfied. For the rest of the analysis we will consider the RHN mass to be 1 TeV. We find, with $y_\ell\sim 10^{-7}$, the lifetime of 1 TeV RHN turns out to be $\tau_N\sim 10^{-12}$ sec, which is much shorter than the typical BBN timescale\footnote{For certain choices of lightest neutrino mass, RHN having such gauge interactions can be long lived enough to give interesting collider signatures like displaced vertices, as studied in~\cite{Das:2019fee}.}. 

\section{Dark matter freeze-in}
\label{sec:dm-pheno}
The interaction strength that controls DM freeze-in production also controls the reaction rate of the new gauge mediator with the SM bath. Therefore, for very small $\gbl$, typically required for freeze-in, the production of $\zbl$ is also heavily suppressed and can be assumed to have a zero initial abundance. It is thus necessary to to consider a two-step production, i.e., to first produce $\zbl$ from the thermal bath and subsequently the DM from the $\zbl$ produced~\cite{PhysRevD.100.095018,PhysRevD.102.035028}. Whether the produced $\zbl$ shall be in thermal equilibrium or not, will depend on the choice of the gauge coupling $\gbl$ that determines the interaction rate between the SM bath and $\zbl$. Thus, there exists a bound on the size of $\gbl$ from these requirements. In order to determine that we need to compare the reaction rate of $\zbl$ production with the Hubble rate $\mathcal{H}$ given by
$\mathcal{H}=\frac{\pi}{3}\,\sqrt{\frac{g_{\star\rho}}{10}}\,\frac{T^2}{\mpl}\,,    $ for a standard radiation dominated Universe, where $g_{\star\rho}$ is the total number of relativistic degrees of freedom contributing to the energy density. Next we note that there exist several channels for the production of $\zbl$ from the thermal bath. All these channels can be broadly classified into three categories based on their time of occurrence as 

\begin{itemize}

\item $v_s=v_d=0$: At this stage both $B-3L_i$ and EW symmetry is exact as none of the scalars in the theory gets a non-zero VEV. As one can see from Eq.~\eqref{eq:ls}, that a non-zero number density of $\zbl$ can still build up from $S\to\zbl\zbl$ decay, however at this stage $\zbl$'s are massless.

\item $v_s\neq 0\,,v_d=0$: Once $B-3L_i$ is broken (but electroweak symmetry is exact), $\zbl$ is massive and can be pair-produced produced both from on-shell decay of the scalar $S\to\zbl\zbl$, along with the $2\to2$ scattering channels: $SS\to\zbl\,\zbl\,,N_R\,N_R\to\zbl\,\zbl\,,f\,f\to\zbl\,\zbl\,,f\,f\to\,W^3(B)\,\zbl$, where $f\in$ SM fermions (chiral) and $V\in\{ W_\mu^3\,,B_\mu\}$ are the SM gauge bosons without the longitudinal degrees of freedom. The SM Higgs doublet is parametrized as $H^T=\bigl\{H^+\,,H^0\bigr\}$, with four propagating massless Goldstone degrees of freedom. Note that, the scalar $S$ can remain in thermal equilibrium with the SM bath through the portal coupling $\lhs$, which can be $\sim\mathcal{O}(1)$ before EWSB. On the other hand, after EWSB, $\lhs$ becomes of the order of $\gbl$ (cf. Eq.~\eqref{eq:cplngs}), however the physical scalar $h_2$ can still maintain thermal equilibrium due to the four point gauge interactions, e.g., $h_2\,h_2\to W^+W^-,\,ZZ$.

\item $v_{s,d}\neq 0$: Once EW symmetry is also broken, then {\it all} states are massive and the SM gauge bosons obtain the longitudinal degrees of freedom. We then obtain two CP-even physical scalars $h_{1,2}$ as elaborated before. Along with this, now the SM fermions are massive and the physical gauge bosons are $W^\pm\,,Z\,,\gamma$.
\end{itemize}

\begin{figure}[htb!]
    \centering
    \includegraphics[scale=0.12]{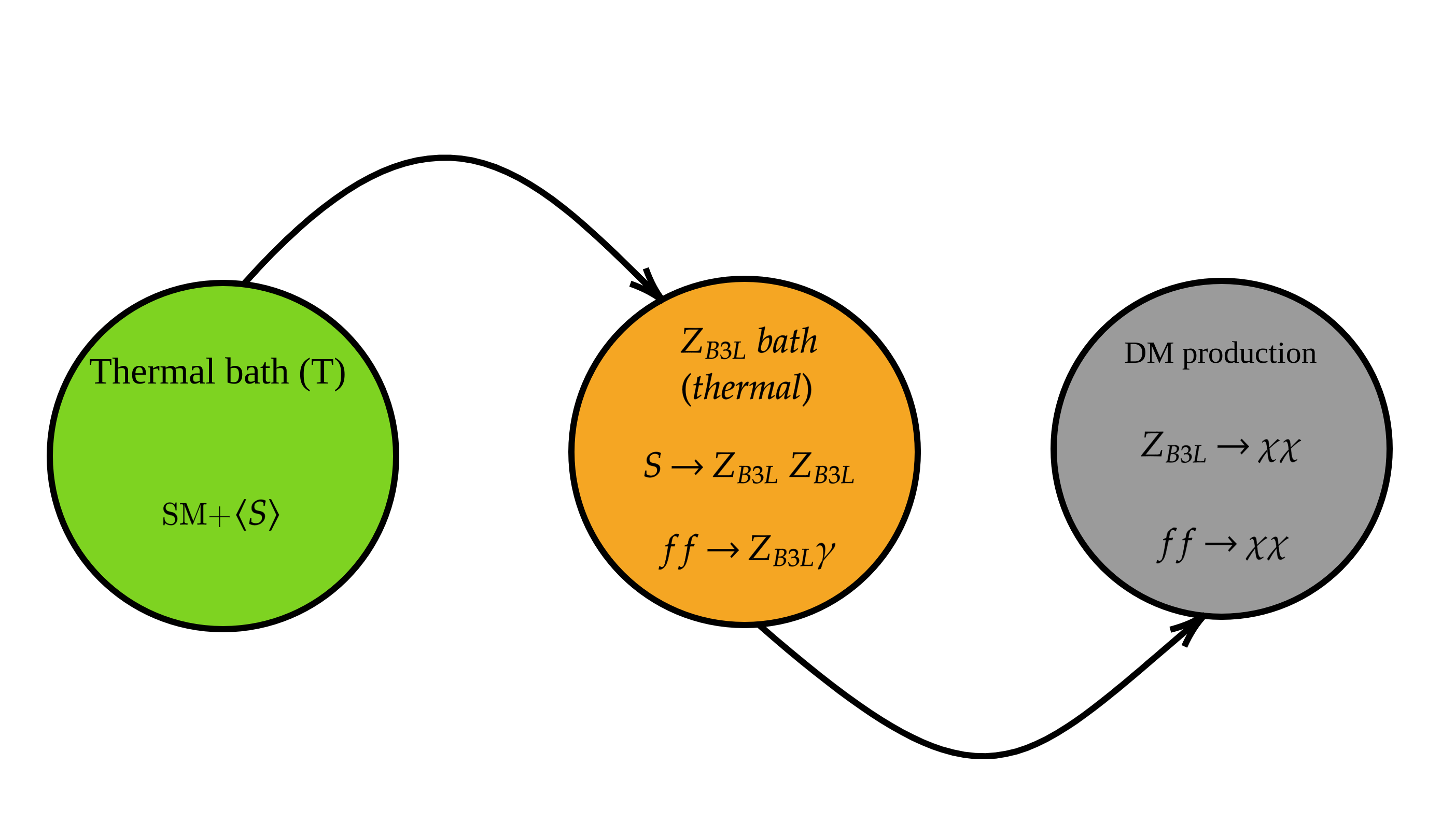}
    \caption{\it DM production via two-step freeze-in.}
    \label{fig:scheme}
\end{figure}

In Fig.~\ref{fig:scheme} we have illustrated the scheme of two-step freeze-in by a cartoon, where we show $\zbl$'s are first thermalized (yellow blob) after being produced from the bath (green blob) and finally the DM yield builds up from the thermalized $\zbl$ density (gray blob).

\subsection{Freeze-in yield}
\label{sec:fi-yld}
As advocated earlier, the computation of DM yield depends on whether $\zbl$ can be considered to be a part of the SM bath. Here we derive the condition under which this is true. We first note {\it before} EWSB the production rate of $\zbl$ is given by
\begin{equation}
 \Gamma_{\text{SM}\to\zbl}^{v_d=0}
    =\begin{cases}
      \Gamma_{S\to\zbl\zbl}\,\frac{K_1(m_S/T)}{K_2(m_S/T)}\,  &\text{for decay}\,,\\[8pt]
     n_{\zbl}\,\langle\sigma v\rangle_{\zbl\zbl} &\text{for scattering}\,,
    \end{cases}
\end{equation}

\begin{figure}[htb!]
    \centering
    \includegraphics[scale=0.32]{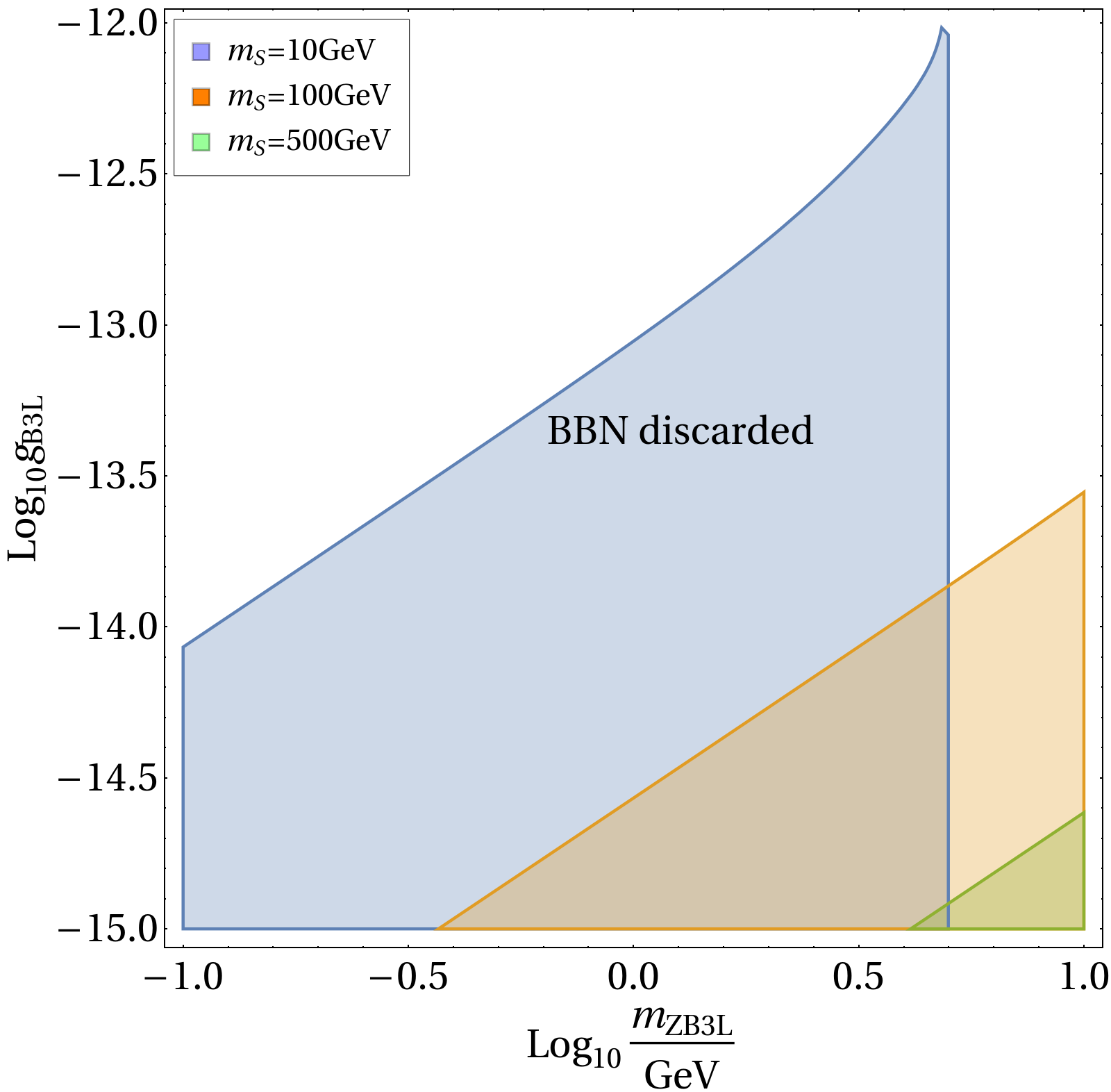}~~~~
    \includegraphics[scale=0.31]{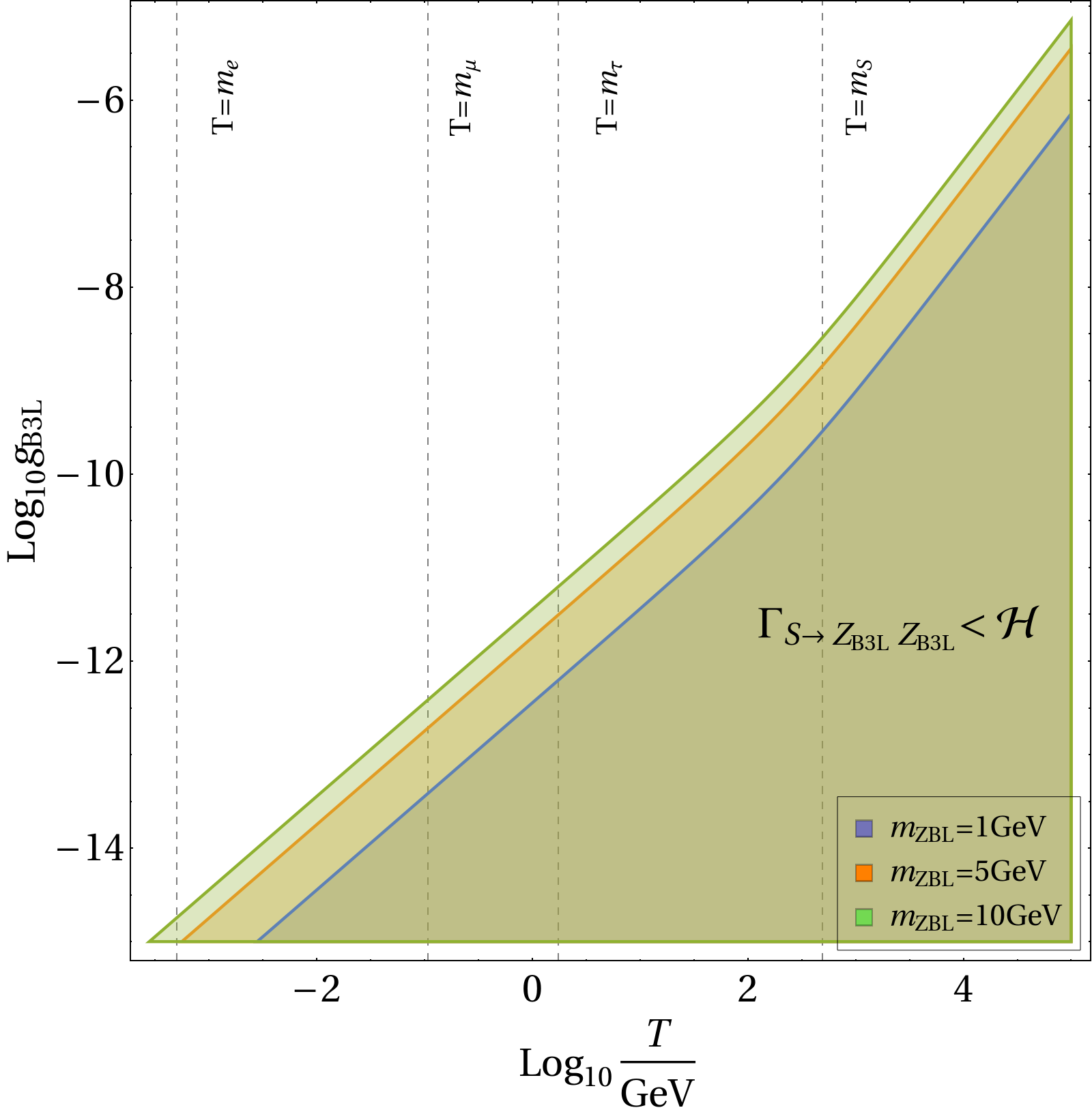}\\
     \includegraphics[scale=0.30]{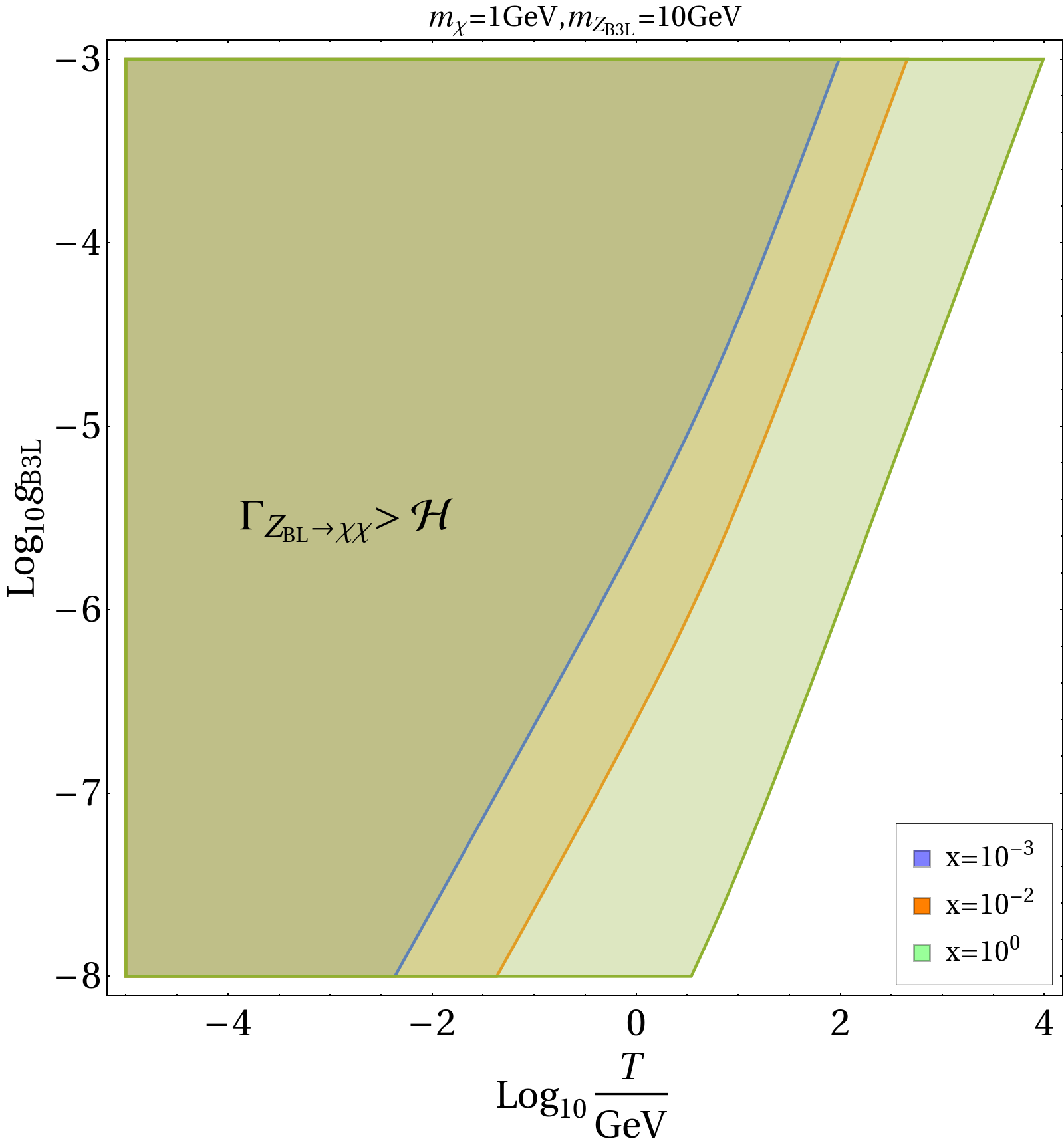}~~~~
     \includegraphics[scale=0.32]{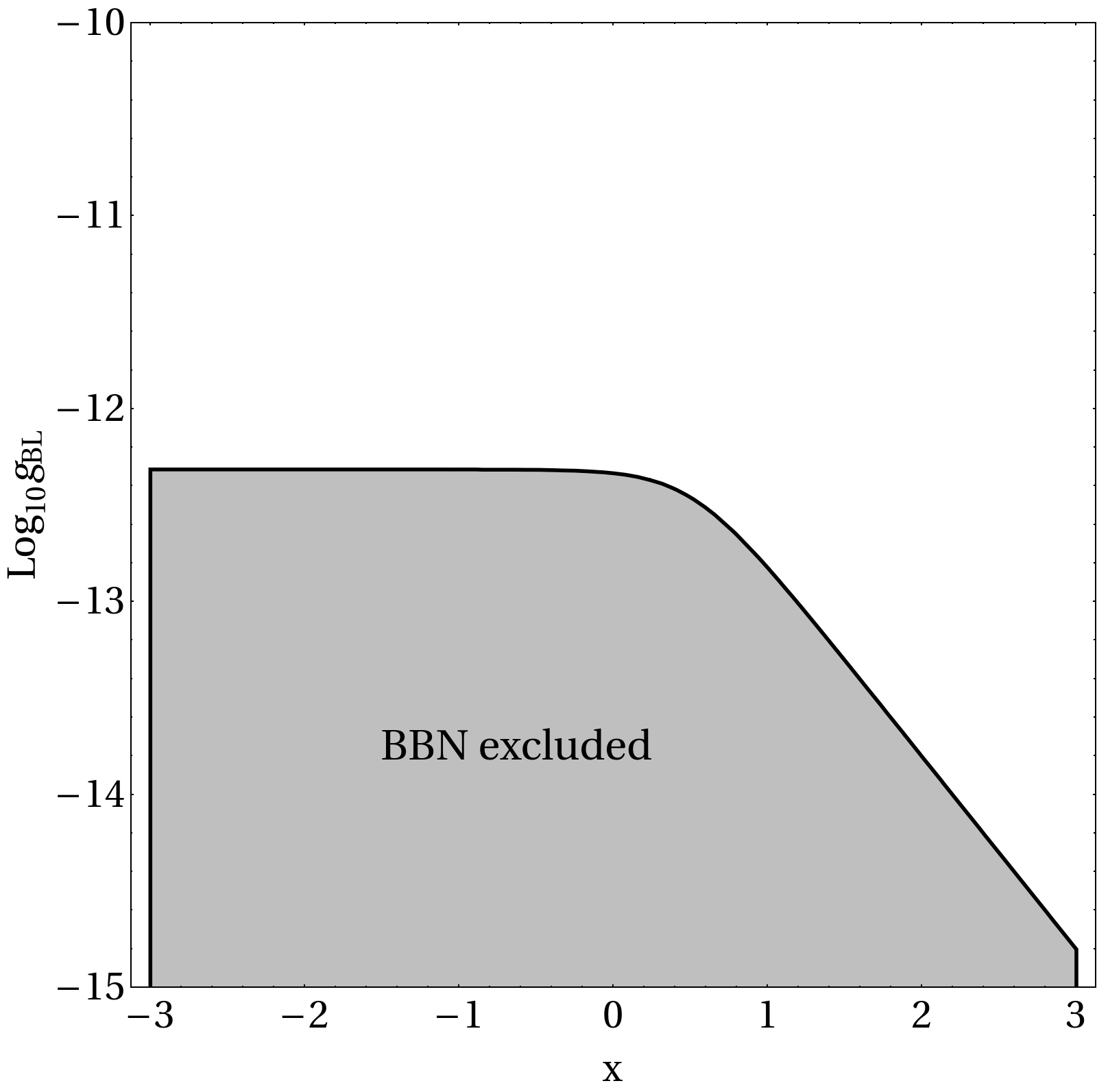}
    \caption{\it Top Left:  Decay rates of $S\to\zbl\zbl$ for different choices of $m_S$ shown via different colors, where the shaded regions are discarded from BBN, requiring $\tau_S \gtrsim 1$ sec. Top Right: The shaded region is where the decay process $S\to\zbl\zbl$ falls out of equilibrium before EWSB. Here different colours correspond to different choices of $\mzbl$. Bottom Left: Same as top right, but comparing DM production rate via decay with the Hubble rate. Bottom Right: The shaded region depicts the region where $\Gamma_{\zbl}^{-1}>1$ sec  (see text).}
     \label{fig:rate}
\end{figure}
\noindent where $n_{\zbl}\sim T^3$ and 
\begin{equation}
\Gamma_{S\to\zbl\zbl}=\frac{9\,\gbl^2}{32
\,\pi\,r^2}\,m_S\,\sqrt{1-4\,r^2}\,(1-4\,r^2+12\,r^4)\,,
\end{equation}
with $r=\mzbl/m_S$. Here $K_i$'s denote the modified Bessel functions. Now, for the $2\to2$ scattering the dominant contribution comes from $f\,f\to W^3(B)\,\zbl$, which has the form $\langle\sigma\,v\rangle\sim\gbl^2\,g_{2(1)}^2/T^2$, while for all other channels we have $\gbl^4$ dependence, and hence negligible given the size of $\gbl$ required for freeze-in. It is important to note here that, there will be an upper bound on the $S$-lifetime from big bang nucleosynthesis (BBN). If the decay occurs after BBN with $\tau_S>1$ sec, entropy production has to be less than $\sim 10\%$ (assuming no change to light element abundances due to it) because the precision measurements of the baryon density through BBN and CMB observations match very well~\cite{Steigman:2012ve,Kaplinghat:2013yxa}. If $S$ decays before BBN, entropy production constraints are non-existent. In the top left panel of Fig.~\ref{fig:rate} we show the BBN bound arising from the lifetime of $S$. Here all the coloured regions are discarded from the requirement of $\tau_S\lesssim 1$ sec. As one can notice, the bound gets weaker for heavier $m_S$ since $\tau_S\propto 1/m_S$. Thus, $m_S=500$ GeV allows $\gbl\gtrsim 10^{-15}$ for $\mzbl\simeq 1$ GeV, as shown by the green region. For $\mzbl\gtrsim 5$ GeV, this turns out to be a bit stronger, requiring $\gbl\gtrsim 10^{-14}$ due to suppression from phase space in the decay width. For the same choice of $m_S$, it is possible to keep $\zbl$ in equilibrium till $T\sim m_S$ for $\mzbl\simeq 1$ GeV with $\gbl\gtrsim 10^{-8}$ as illustrated in the top right panel (blue contour). For heavier $\zbl$, e.g., $\mzbl=10$ GeV (orange contour) this bound slightly changes. Since we are interested in $\mzbl\lesssim 10$ GeV, therefore it is possible to keep $\zbl$ in thermal equilibrium with $\gbl\gtrsim 10^{-8}$ that also satisfies the BBN bound for $m_S=500$ GeV. Once the electroweak symmetry is also broken $(v_d\neq0)$, $\zbl$ can be produced via the decay $h_{1(2)}\to\zbl\zbl$ (if kinematically allowed) and also via the dominant scattering channel $f\,f\to\zbl\,\gamma$. The full expressions of all the decay and scattering processes contributing to $\zbl$ production in the post-EWSB scenario are gathered in Appendix.~\ref{sec:app-zbl-prod}. In this regime we obtain similar bounds on $\gbl$ for $\zbl\lesssim 10$ GeV, with the masses and mixing of our choice as mentioned earlier.
\begin{figure}[htb!]
    \centering
    \includegraphics[scale=0.42]{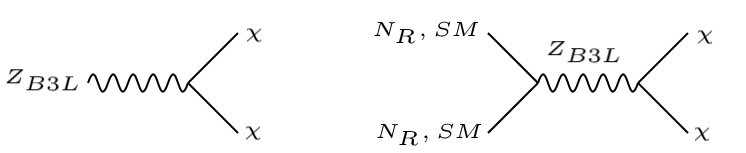}
    \caption{\it DM production channels from on-shell decay of $\zbl$ (left) and scattering mediated by $\zbl$ (right).}
    \label{fig:feyn-diag}
\end{figure}
It is necessary to ensure that the DM is produced out of equilibrium with the choices of $\gbl$ obtained above. This is depicted in the bottom left panel of Fig.~\ref{fig:rate}. Here we show the bound on $\gbl$ and $x$ requiring $\langle\Gamma_{\zbl\to\chi\chi}\rangle<\mathcal{H}$, where we consider the decay to be dominant DM production channel. We find, with $x=10^{-3}$, it is possible to achieve the condition of non-thermalization till $T\sim 1$ MeV. Larger values of $x$ certainly makes the the DM thermal at a comparatively larger temperature simply because $\Gamma_{\zbl\to\chi\chi}\propto x^2\,\gbl^2$. However, the decay reaction rate can not compete with the Hubble rate at a very high temperature as it is evident from the bottom left panel of Fig.~\ref{fig:rate}. For DM production via annihilation, this condition however is trivially satisfied since $\langle\sigma v\rangle\propto x^2\,\gbl^4$. Finally, we would like to mention that the lifetime of $\zbl$ turns out to be $\tau_{\zbl}\ll\tau_\text{BBN}$ for $\gbl\gtrsim 10^{-12}$ as shown in the bottom right panel, taking all decay channels of $\zbl$ into account.

\subsubsection*{Boltzmann equation for dark matter yield}
As it is usually done, while considering the production of DM from any SM/BSM particle, the latter is implicitly assumed to be in thermal equilibrium. Hence we usually do not need to solve a system of coupled Boltzmann equations, since the equilibrium number density is assumed for the decaying mother particle. But, here due to very low interaction strength of $\zbl$ (due to small $\gbl$ which also a universal coupling), it will not be in thermal equilibrium with the rest of the particles. So, first, we find the comoving number density of $\zbl$ by solving its Boltzmann equation. Then we use this to find the relic density of the DM. Thus the DM production in the present scenario is a two-step process (cf. Fig.~\ref{fig:scheme}). Once a non-zero number density of $\zbl$ is formed from the SM bath, it is then possible to produce DM from the on-shell decay of $\zbl$ and/or scattering of the bath particles mediated by $\zbl$, depending on the mass of the DM. The corresponding Feynman graphs are shown in Fig.~\ref{fig:feyn-diag}. Note that, the decay channel is always present irrespective of the EWSB (as long as kinematically accessible), while all the SM particles are assumed to be massless above the EWSB scale and massive below. Note that, in the case $v_s=v_d=0$, since the $\zbl$ is massless, hence the DM can be produced only through $s$-channel $Z'$ mediation (cf. Fig.~\ref{fig:feyn-diag}). However, because of $1/s\sim1/T^2$ dependence of the cross-section, the contribution to DM yield from this channel will be heavily suppressed since $T\gg v_s$. Since it is evident that the evolution of DM number density is related to the primordial number density of $\zbl$, hence one has to solve a set of coupled Boltzmann equations (cBEQ) to capture the evolution of DM number density properly. In Eq.~\eqref{eq:cBEQ} we have written down the cBEQ in terms of the dimensionless quantity $z=m_\chi/T$ that acts as a proxy for the temperature of the thermal bath. The first equation takes care of the evolution of $\zbl$ yield, where we define the yield as the ratio of number density to entropy density of the visible sector as $Y_i=n_i/s$ for the $i^\text{th}$ particle concerned. The equilibrium yield is given by $$Y_\text{eq}=\frac{45}{4\,\pi^4}\,\frac{g_i}{g_{\star s}}\,(m_i/T)^2\,K_2(m_i/T)\,,$$ where $g_i$ is the intrinsic number of degrees of freedom of the species concerned (which is unity for $S$) and $g_{\star s}$ is the total number of relativistic degrees of freedom present in the thermal bath. Note that, depending on the heaviside $\Theta$ function, $\zbl$ production channels before and after EWSB shall appear. The second equation takes care of the evolution of DM yield, where we consider contribution to the DM number density both from $\zbl$ decay (first term) and scattering mediated by $\zbl$ (second term). We solve this set of equations numerically to obtain the final DM relic abundance, utilizing the DM production cross-sections and decay reported in Appendix.~\ref{sec:app-decay-ann}. 

\begin{eqnarray}
    \frac{dY_{\zbl}}{dz} &=& \frac{s(m_\chi)}{\mathcal{H}(m_\chi)} \frac{1}{z^2}\langle \sigma v_{_{ff\to\zbl\gamma}} \rangle \big( {Y_{\rm eq}}^2-Y_{\zbl} Y_{\rm eq}\big) - \frac{z}{\mathcal{H}(m_\chi)}\langle \Gamma_{\zbl \to \chi\,\chi }\rangle Y_{\zbl} \nonumber \\
    && +\frac{z}{\mathcal{H}(m_{\chi})} \Big( \langle \Gamma_{S \to \zbl \zbl }  \rangle \Theta[z_{\rm ew}-z] \big(Y_{S}^\text{eq}-Y_{\zbl}\big)  \nonumber \\ 
    &&    ~~~~~~~~~~~~~~~~~~~~~~~~~~~~ +\langle \Gamma_{h_{1,2} \to \zbl \zbl }  \rangle \Theta[z-z_{\rm ew}] \big( Y_{h_{1,2}}^\text{eq} - Y_{\zbl}\big) \Big) \nonumber  \\
    \frac{dY_\chi}{dz} 
    &=&   \frac{z}{\mathcal{H}(m_\chi)}  \langle \Gamma_{\zbl \to \chi~\chi}\rangle Y_{\zbl} + \frac{s(\mdm)}{\mathcal{H}(\mdm)}\, \frac{1}{z^2}\,\langle\sigma v_{\text{SM}\,\text{SM}\to\chi\chi}\rangle\,Y_\text{eq}^2\, , 
    \label{eq:cBEQ}
\end{eqnarray}
\begin{figure}[htb!]
    \centering
    \includegraphics[scale=0.5]{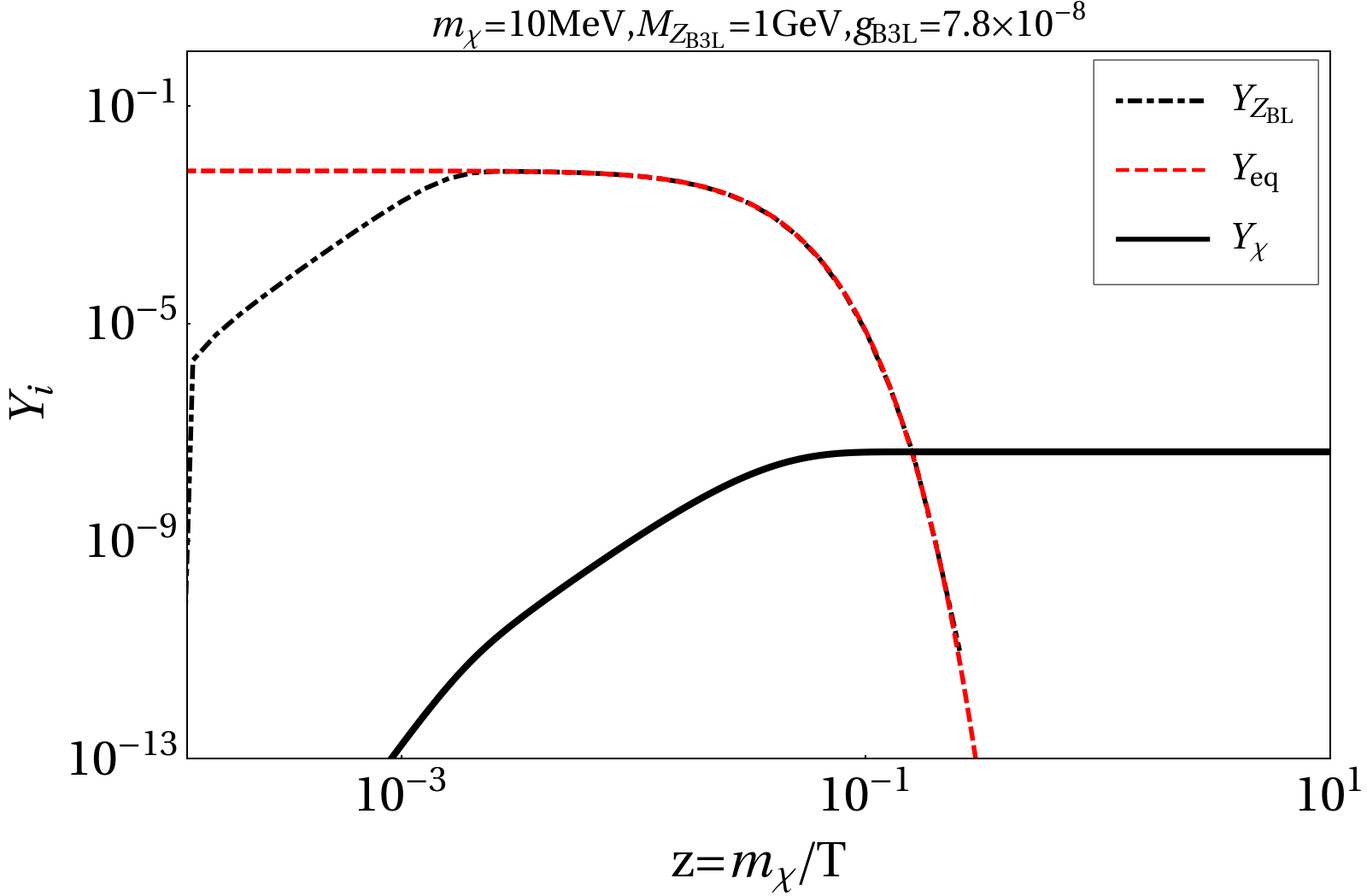}
    \caption{\it Evolution of yields $Y_i$ with inverse temperature $z$ ($=\frac{m_\chi}{T}$). Here we have fixed $x=10^{-3}$ such that right relic abundance is satisfied by the asymptotic DM yield shown via the thick black curve.}
    \label{fig:yld}
\end{figure}
where $s(m_\chi)=\frac{2\pi^2}{45}\,g_{\star s}\, m_\chi^3$ and $\mathcal{H}(T=m_\chi)=\sqrt{\frac{\pi^2 g_{\star\rho}}{90}}\, \frac{m_\chi^2}{M_{\rm pl}}$. In order to visualize the evolution of the yields with the dimensionless parameter $z$, we plot $Y_{\chi\,,\zbl}$, together with $Y_\text{eq}$ in Fig.~\ref{fig:yld}. Here we see, as expected, $\zbl$ is first produced from the thermal bath via decay and scattering of the bath particles, as denoted by the black dot-dashed curve. Once it is produced, it starts following the equilibrium distribution from $z\sim 10^{-3}$ onward. As the $\zbl\to\chi\chi$ decay is completed, the yield $Y_\chi$ saturates around $z\sim 10^{-1}$ and remains fixed till today, while the $\zbl$ abundance keeps tracking the equilibrium. The final DM relic density is determined using
\begin{equation}
\Omega_\chi\,h^2 = \left(2.75\times 10^8\right) \left(\frac{m_\chi}{\text{GeV}}\right) Y_\chi(T_0)
\label{eq:relicX}    
\end{equation}
where $T_0$ is the temperature of the Universe at the present epoch. We must also remember that the  PLANCK~\cite{PLANCK:2018vyg} allowed relic density allows: $\Omega_\text{DM} h^2 = 0.11933\pm 0.00091\,,$ which we will use to constrain the relic density allowed parameter space. 

\subsection{Viable parameter space}
\label{sec: paramspace}
Here we illustrate the relic density allowed parameter space in $\gbl-\mzbl$ plane for some benchmark choices of the DM mass, and for different choices of DM charge $x$. This is also the experimental plane which is of our present interest. Thus, we are going to project both already existing and future bounds from different experiments on this plane, together with contours satisfying observed DM abundance. We fix two benchmark values of the DM mass: 10 MeV and 6 GeV. In the former case the dominant contribution to the DM relic abundance comes from on-shell decay of $\zbl$, together with sub-dominant contribution from SM scattering. In the later case, on the other hand, the entire contribution arises from scattering since the decay contribution is completely switched off as $\mzbl< 2\,\mdm$. The black tilted straight line contours in Fig.~\ref{fig:relic1} correspond to different choices of $x$ satisfying the observed relic abundance. Here we see, larger $x$ is capable of producing the correct relic for smaller $\gbl$, while the bound on $\gbl$ is sufficiently relaxed for smaller $x$. Since decay plays the dominant role in setting the right relic abundance in this case, the DM yield can be analytically expressed as
\begin{equation}
Y_\chi \approx \frac{135}{g_{\star s}\,\sqrt{g_{\star\rho}}}\,\frac{x^2\,\gbl^2}{4\,\pi^5}\,\frac{\mpl\,\mzbl^3}{\mdm^4}\,\frac{\sqrt{1-4\,r^2}\,(1+2\,r^2)}{r^4}\,,
\end{equation}
with $r=\mzbl/m_\chi$. This shows, for a fixed DM mass $\mdm$ and fixed $x$, in order to satisfy the PLANCK data, one has to increase $\gbl$ with the increase in $\mzbl$ since $Y_\chi\propto x^2\,\gbl^2/\mzbl$. This pattern follows in Fig.~\ref{fig:relic1}. On the other hand, for $\mdm>\mzbl/2$, the only contribution to DM yield arises from 2-to-2 scattering processes mediated by $\zbl$. The DM yield in that case can be approximated as (in the limit of massless SM states)
\begin{equation}
\frac{dY_\chi}{dz}\approx\frac{\gbl^4\,x^2}{\pi^8\,\sqrt{g_{\star\rho}}\,g_{\star s}}\,\frac{\mpl}{\mdm}\Biggl[4\,K_1(z)^2+z\,\sqrt{\pi}\,G_{1,3}^{3,0}\left(z^2\Bigg|
\begin{array}{c}
 1 \\
 -\frac{1}{2},-\frac{1}{2},\frac{1}{2} \\
\end{array}
\right)\Biggr]\,,
\label{eq:yld-ann}
\end{equation}
where $G^{m,n}_{p,q}$ are the Meijer's $G$ functions, and in the case of freeze-in with $\mdm/T_\text{FI}\simeq\mathcal{O}(1)$ at the time of freeze-in, the numerical value turns out to be $\mathcal{O}(10^{-1})$. Due to $\gbl^4$ dependence, in order to satisfy the relic abundance, one has to choose a larger $x$ for a given $\gbl$ compared to the case where DM production takes place via decay. Note that, the expression in Eq.~\eqref{eq:yld-ann} is independent of the mediator mass. This is exactly what is reflected in Fig.~\ref{fig:relic2}, where we see the straight line contours of right relic abundance, shown via different colours, do not depend on $\mzbl$.

\subsubsection*{Constraints on the dark matter parameter space}
Since the SM fermions and RHN channels are always present irrespective of the $B-3L_i$ gauge choices, the relic density allowed parameter space remains almost unchanged for different choices of the $U(1)_{B-3L_i}$ gauge symmetry. The main constraint on $\gbl$ and $\mzbl$ arise from the search of $U(1)_{B-3L_i}$ gauge boson in different collider and beam dump experimental facilities, together with constraints from BBN on the lifetime of $\zbl$ and also from non-standard neutrino interactions (NSI) between neutrinos and matter (for recent review, see, for example~\cite{Farzan:2017xzy, Proceedings:2019qno}). Apart from BBN, at low masses and small couplings, there is also a very strong bound from Supernova 1987A (SN1987A)~\cite{Chang:2016ntp, Knapen:2017xzo, Croon:2020lrf, Dev:2021qjj}, which occurs due to the presence of the weakly coupled vector boson that can significantly contribute to the energy loss during supernova explosion, thereby shortening the duration of the observable neutrino pulse emitted during core collapse. We mostly follow the constraints derived in Ref.~\cite{Bauer:2020itv}, where the authors have performed an updated analysis, and contrast current constraints and the sensitivity reach of proposed experiments for $U(1)_{B-3L_i}$ gauge boson for neutrino observables and quark flavour transitions with beam dump, collider and precision experiments. As we will see, depending on the choice of the gauge symmetry, the relic density allowed parameter space for the DM can become more constrained or comparatively relaxed.

\begin{figure}[htb!]
$$
\includegraphics[scale=0.37]{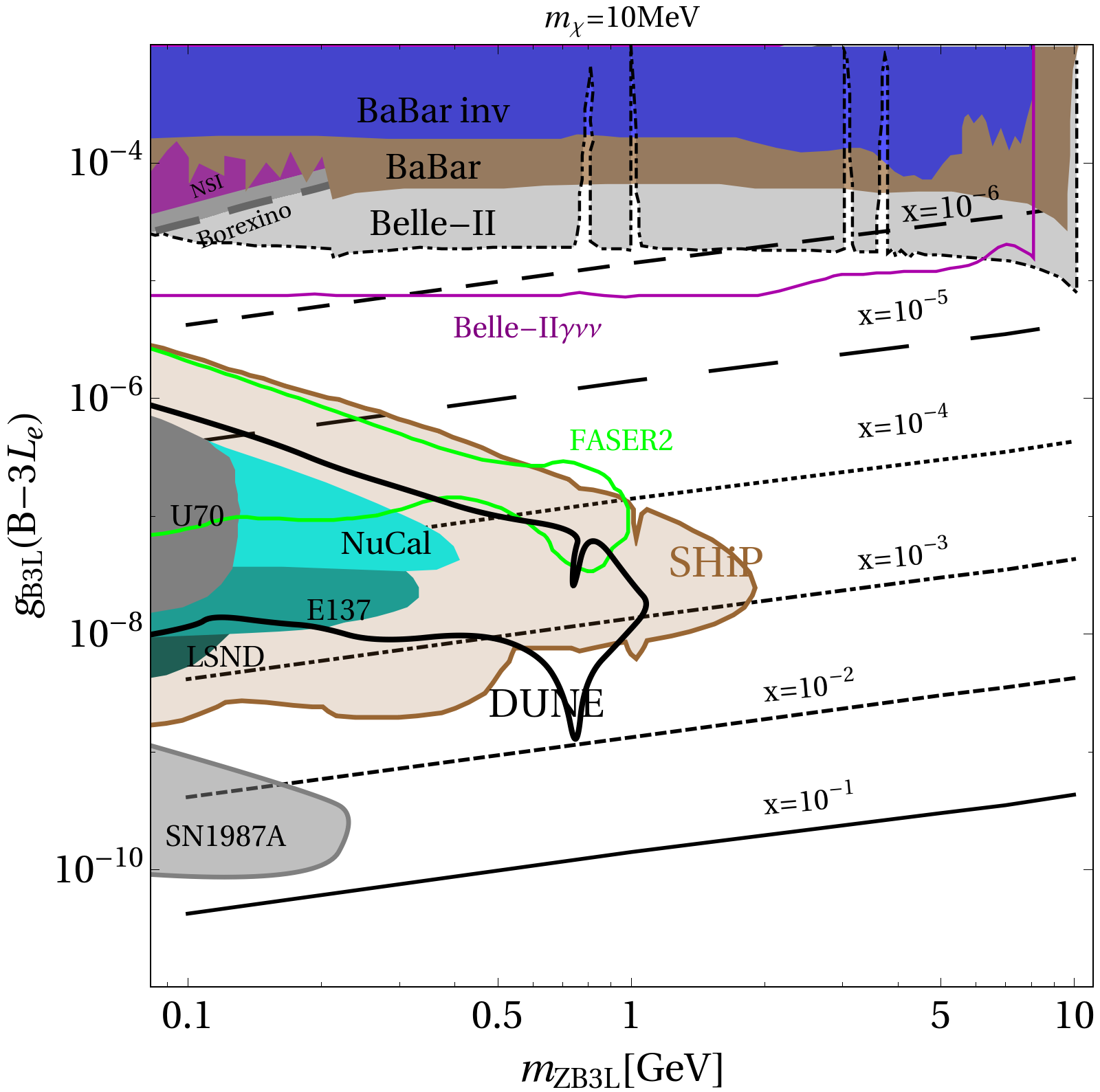}~~~\includegraphics[scale=0.37]{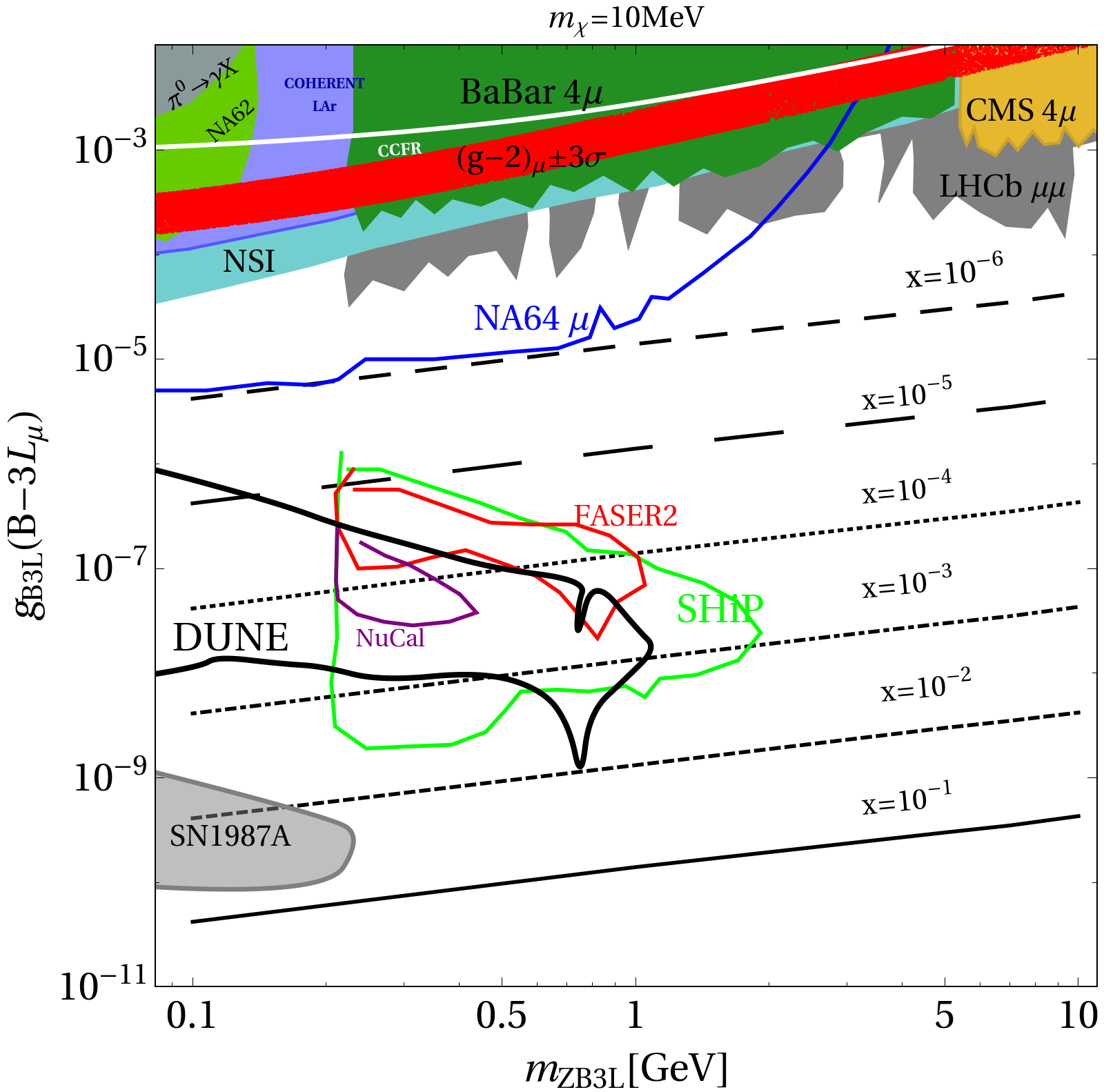}
$$
$$
\includegraphics[scale=0.37]{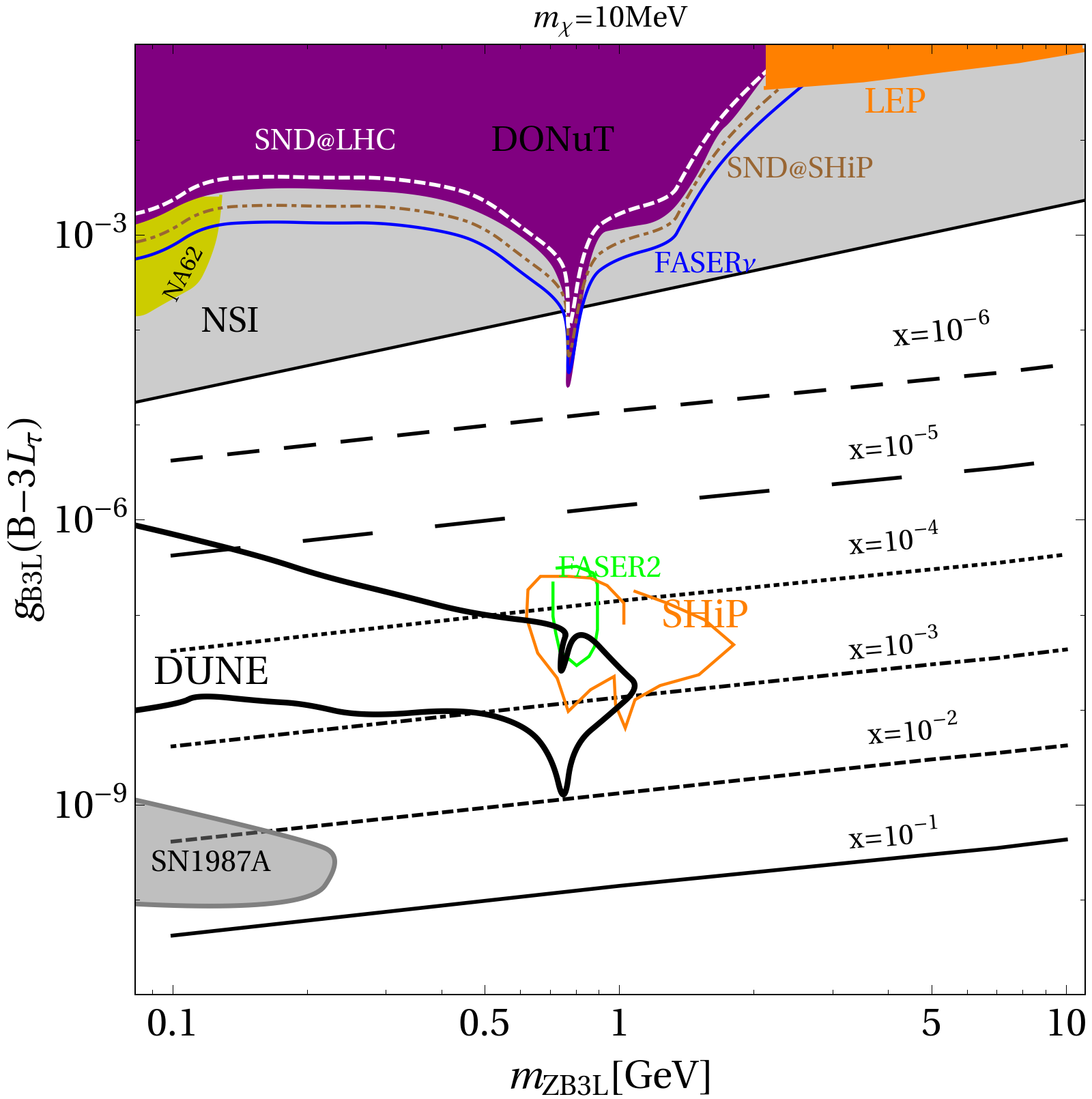}
$$
    \caption{\it The black tilted straight contours having different patterns satisfy PLANCK observed relic density for different choices of the DM charge $x$, in increasing order from top to bottom. The DM mass is fixed at 10 MeV. Bounds from different experiments are shown in different colours.}
    \label{fig:relic1}
\end{figure}

\begin{figure}[htb!]
$$
\includegraphics[scale=0.37]{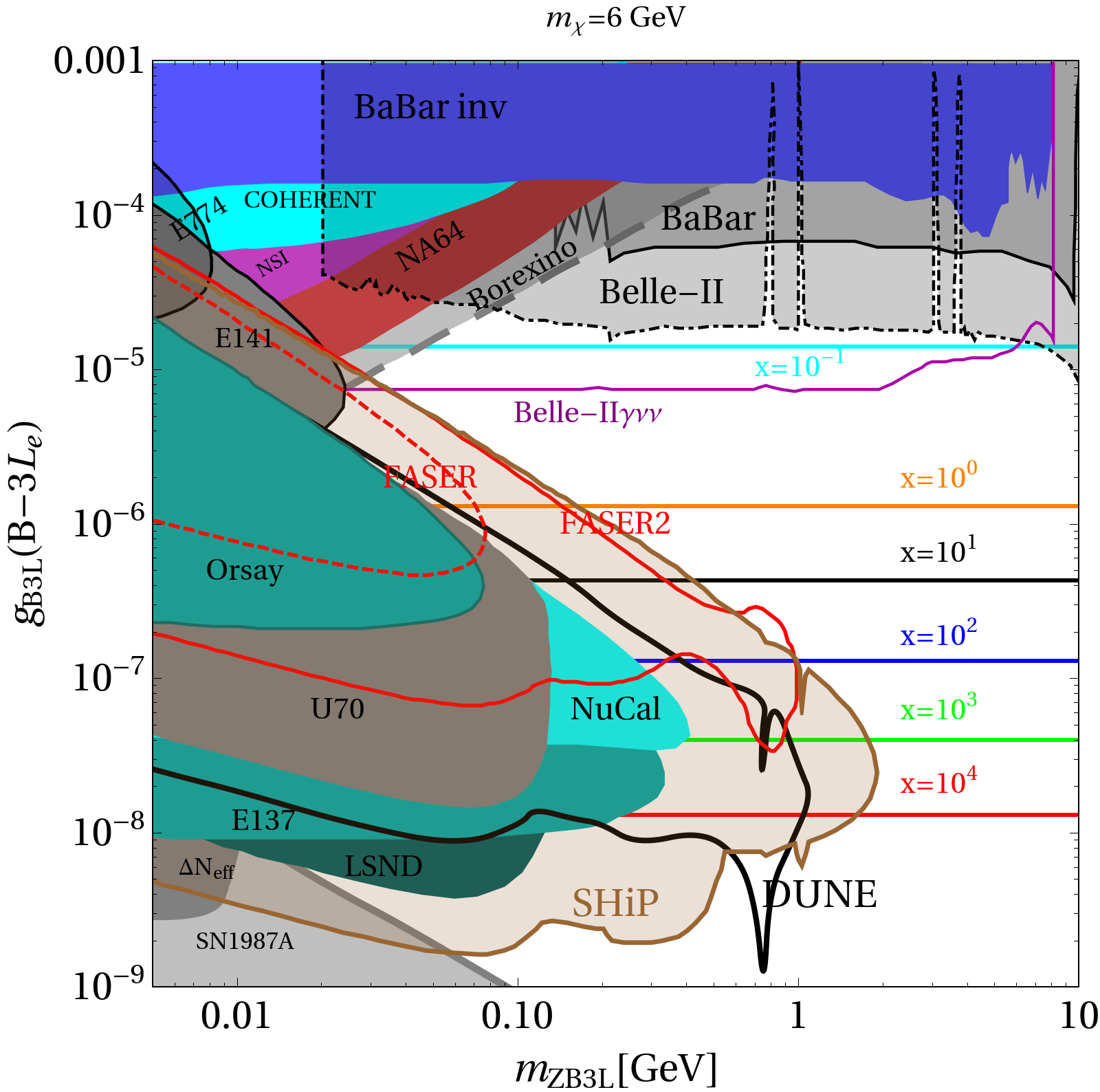}~~~\includegraphics[scale=0.37]{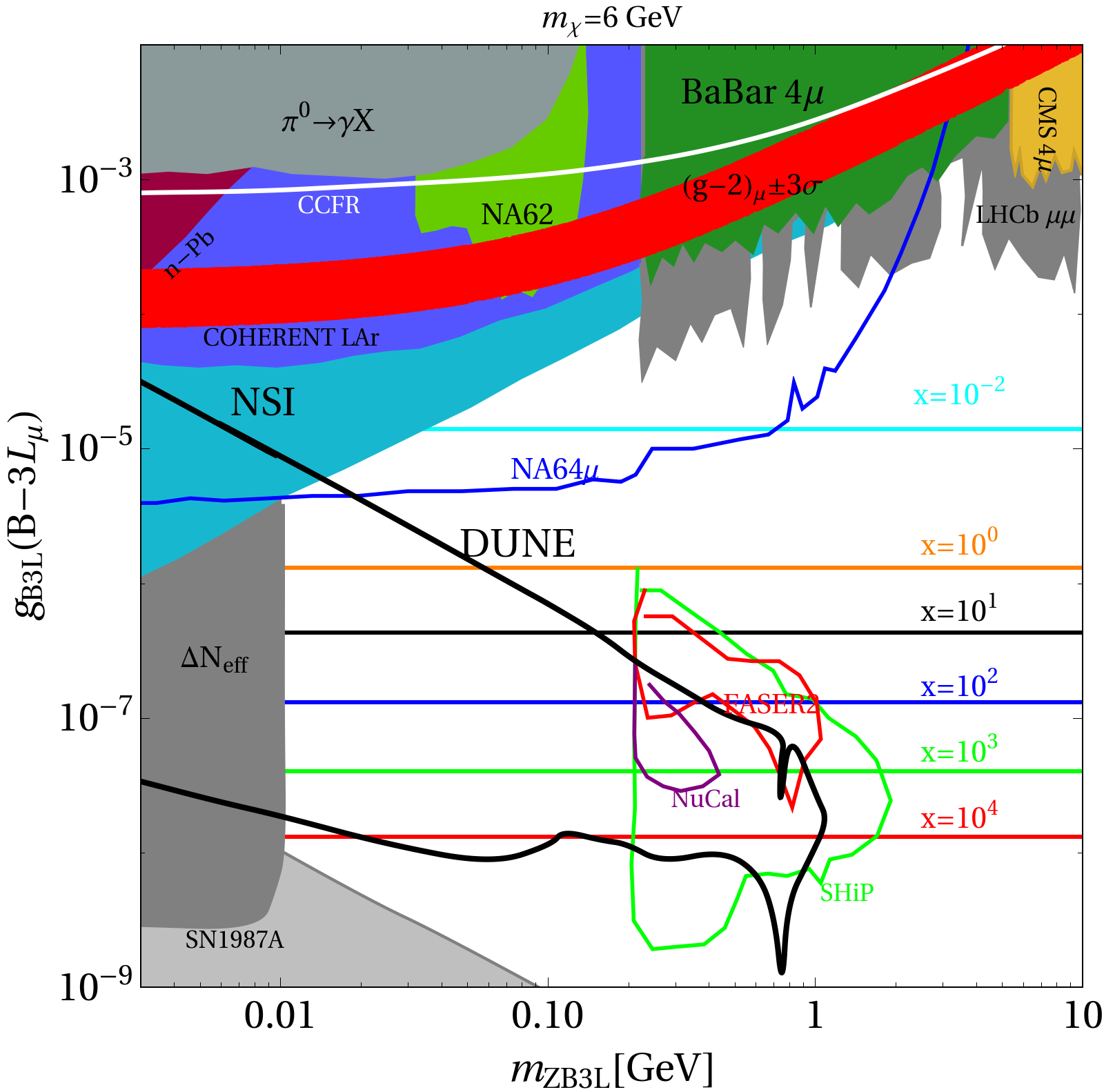}
$$
$$
\includegraphics[scale=0.37]{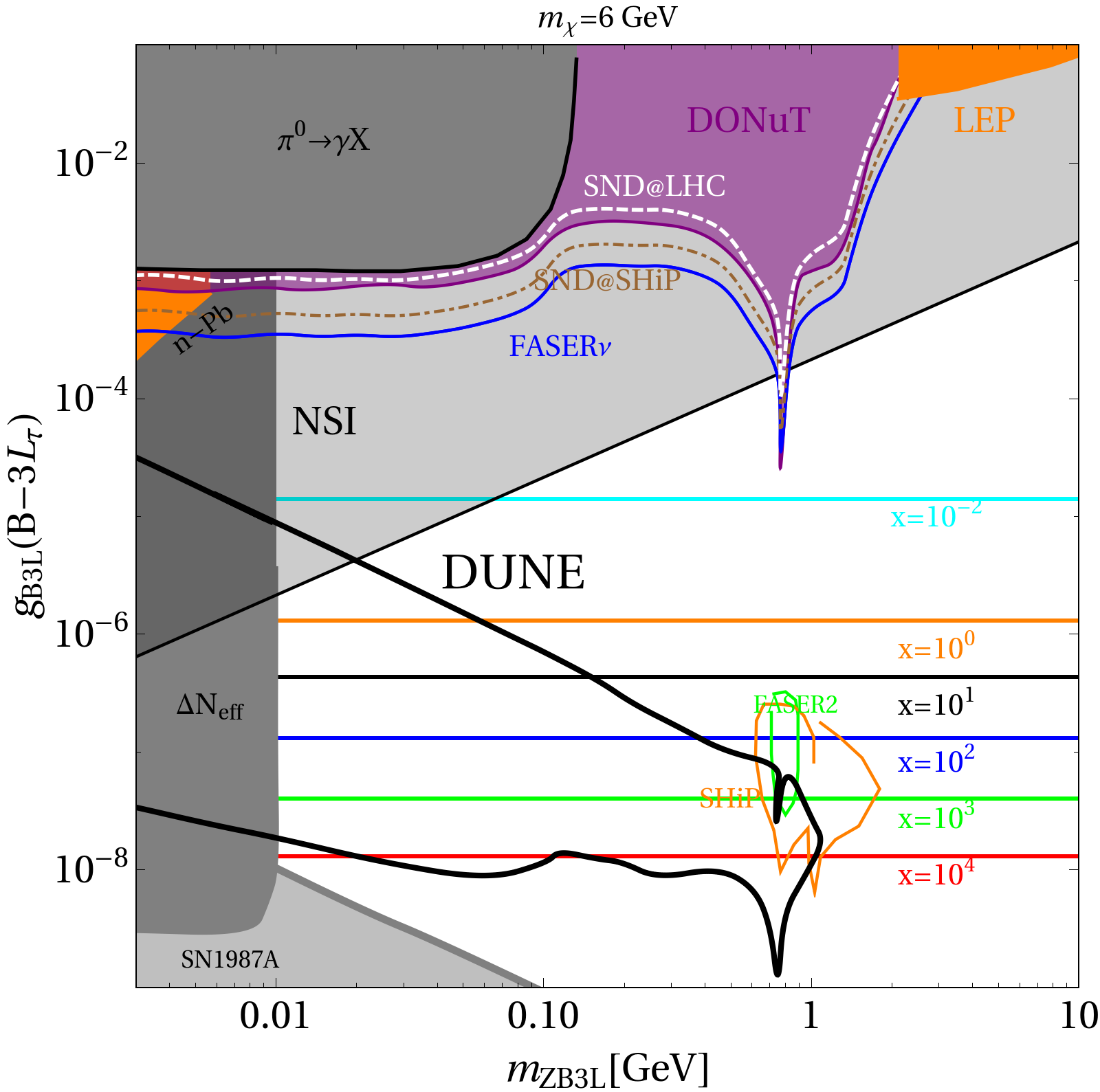}
$$
    \caption{\it Same as Fig.~\ref{fig:relic1}, but with DM mass $\mdm=6$ GeV. Here the straight lines with different colors denote right relic abundance for different choices of $x$ in decreasing order from bottom to top (see text).}\label{fig:relic2}
\end{figure}

As it is evident from Fig.~\ref{fig:relic1} and Fig.~\ref{fig:relic2}, larger values of $x$, i.e., smaller $\gbl\lesssim 10^{-9}$ are typically unconstrained from present as well as future experimental sensitivities. For DM mass of 10 MeV, this corresponds to $x\gtrsim 10^{-2}$, while for a DM of mass 6 GeV this is true for $x\gtrsim 10^5$, due to the absence of decay. However, all such $\gbl$ values can still produce the observed relic abundance. In case of $U(1)_{B-3L_e}$, with $x\gtrsim 10^{-3}$, corresponding to $\gbl\gtrsim 10^{-8}$, searches from beam dump and fixed target experiments like E137, E141, E774~\cite{Bjorken:2009mm}, NuCal/U70~\cite{Blumlein:2011mv, Blumlein:2013cua} , LSND~\cite{Essig:2010gu} already discard $\mzbl\lesssim 200$ MeV. Future novel fixed target experiments with high intensity beams, e.g., SHiP~\cite{SHiP:2015vad} or the LHC forward detector FASER~\cite{Feng:2017vli, FASER:2018eoc, FASER:2018bac, FASER:2019aik}, can provide a much better sensitivity in probing the relic density allowed parameter space for $\mzbl\sim 1.8$ GeV. Increasing $\gbl$ further makes the viable parameter space available to $e^+e^-$ colliders like Belle-II~\cite{Belle-II:2018jsg}, which is found to be more sensitive than the existing limits from BaBar~\cite{BaBar:2014zli, BaBar:2017tiz}. For $\mzbl\lesssim 0.1$ GeV, there arise bounds due to elastic neutrino-nucleus scattering (CE$\nu$NS) on liquid argon, measured at the COHERENT~\cite{COHERENT:2020iec} facility. Bounds from the neutrino non-standard interactions (NSI)~\cite{Coloma:2015pih, Coloma:2020gfv} and Borexino~\cite{Amaral:2020tga, Bellini:2011rx, Borexino:2017rsf} can constraint the parameter space for $\gbl\gtrsim 10^{-5}$. We have adopted the NSI limits derived in~\cite{Coloma:2020gfv}, where the authors have considered lepton flavor dependent $U(1)'$ interactions\footnote{For similar limits from NSI see Ref.~\cite{Heeck:2018nzc}.}. For  $\mzbl\lesssim 10$ MeV, the $\Delta N_\text{eff}$ bound due to heating of the neutrino bath during neutrino decoupling in the early Universe~\cite{Kamada:2015era, Escudero:2019gzq}, becomes more stringent, which is substantially weaker for $\mzbl>10$ MeV (cf.Fig.~\ref{fig:relic2}). 

Measurements of muonic neutrino tridents: $\nu_\mu\to\nu_\mu\,\mu^+\,\mu^-$ performed at the CCFR~\cite{PhysRevLett.66.3117, Altmannshofer:2019zhy} facility has found no excess over the SM expectations, thus posing a bound on $\gbl-\mzbl$ plane in the case of $U(1)_{B-3L_\mu}$, as shown by the solid white curve in the top right panel of Fig.~\ref{fig:relic1} and Fig.~\ref{fig:relic2}. Similar to the $B-3L_e$ scenario, in $B-3L_\mu$ case, future sensitivity of FASER2 and SHiP can provide extremely good probe to the relic density allowed parameter space for $\gbl\lesssim 10^{-6}$. Importantly, DM parameter space with $\gbl\gtrsim 10^{-5}$ can be probed with the planned upgraded run of NA64 with a dedicated muon beam (shown by the blue curve). Again, bounds from NSI and COHERENT becomes important for $\mzbl\lesssim 0.1$ GeV. Here we also show the 3$\sigma$ preferred region of the observed anomalous $(g-2)_\mu$ excess in red (See Appendix~\ref{sec:muong2} for the computation) \cite{Davier:2017zfy,Davier:2019can,Muong-2:2006rrc,Aoyama:2020ynm,Muong-2:2021ojo}. However, a common explanation of both $(g-2)_\mu$ anomaly and DM abundance is possible with $\gbl\sim 10^{-3}$, for which one has to tune the $x$-charge accordingly to ensure freeze-in production remains non-thermal. 

Due to hadronic final state, the parameter space for $U(1)_{B-3L_\tau}$ scenario is comparatively relaxed than the other two cases. The most stringent bound comes from NSI-induced neutrino oscillations that typically constraints $\gbl\gtrsim 10^{-5}$. Large $\gbl$ values are constrained from the measurement of scattering of tau neutrinos produced in the decay of $\zbl$ with DONuT detector~\cite{DONuT:2007bsg}\footnote{$\nu_\tau$ related NSI and decays can also become important if DUNE experiment involves modified beam flux as shown in Ref.~\cite{Ghoshal:2019pab,Ghoshal:2020hyo}.}. LEP measurement of partial width of $Z$ into taus can also constraint a part of the parameter space for large $\gbl$~\cite{Ma:1998dp} (shown in orange in the bottom panel of Fig.~\ref{fig:relic1} and Fig.~\ref{fig:relic2}). We also project bounds from the proposed scattering and neutrino detector by SHiP, known as SND@SHiP (white dashed curve)~\cite{SHiP:2015vad} and also a rather weaker bound from SND@LHC (brown dot-dashed curve)~\cite{SHiP:2020sos}. Although the SND@SHiP
proposal benefits from a much larger neutrino flux, its sensitivity is limited by systematic uncertainties, resulting in roughly the same reach as FASER$\nu$ (blue solid curve)~\cite{Kling:2020iar}. The reach of SND@SHiP can be further improved by reducing the flux uncertainty as proposed by the DsTau experiment~\cite{Aoki:2017spj, DsTau:2019wjb} at the CERN-SPS, that plans to study tau-neutrino production by directly measuring $D_s\to\nu_\tau\,X$ decay. The search for a light $Z'$ boson at high-intensity facilities such as the near
detector complex of the Deep Underground Neutrino Experiment (DUNE)~\cite{DUNE:2015lol, DUNE:2016hlj} has been studied in~\cite{Berryman:2019dme, Dev:2021qjj}. Here we see, DUNE can probe heavier $\zbl$ and smaller $\gbl$, covering most of the region of existing beam-dump searches. The DUNE loses its sensitivity in the visible channel for $\mzbl<2\,m_e$, where the dominant decay turns out to be into light neutrino pairs~\cite{Bakhti:2018avv}. The relic density allowed parameter space for $\gbl\lesssim 10^{-6}$ is totally unconstrained from experimental bounds for $\mzbl\gtrsim 10^{-2}$ GeV for the $U(1)_{B-3L_\tau}$ scenario. However, future projections from FASER2 and SHiP can potentially probe a part of the parameter space for $\mzbl$ around 1 GeV. It is interesting to note that experiments like SHiP and FASER can probe different mass regime of $\zbl$, hence a different DM parameter space, depending on the flavour structure. For $B-3L_\mu$ and $B-3L_\tau$ future projections from SHiP and FASER cover mostly higher $\mzbl$ than in $B-3L_e$. Finally, we note that the $\zbl$ mass also lies within the window of the new ``X17" boson of mass $17$ MeV, observed by the ATOMKI collaboration~\cite{PhysRevLett.116.042501, Feng:2016jff} in the decay of excited states of Beryllium yielding $e^+\,e^-$ pairs \footnote{Such scenarios are also being searched for in positron beam-dump experiment like PADME \cite{Nardi:2018cxi}.}.

In the absence of kinetic mixing between $U(1)_Y$ and $U(1)_X$ gauge groups, the models under consideration will contribute to flavour changing quark transitions at one loop level. As a result, the light $\zbl$ can be produced in meson decays and subsequently decay to leptons to which it couples to. Hence constraints from FCNC decays such as $B \to K^{(*)} \ell^+ \ell^-$, $B^+ \to K^+ \nu \bar{\nu}$, $B \to \pi + \text{invisible}$, $K \to \pi + \text{invisible}$ etc. can, in general, put strong bounds on the $\{\gbl,\mzbl\}$ parameter space. For example, LHCb has put upper limits on the branching fractions of $\mathcal{B}(B^{0} \to K^{0*} Z' ) \times \mathcal{B}(Z'  \to \mu^+ \mu^-)$ \cite{LHCb:2015nkv} and $\mathcal{B}(B^{+} \to K^{+} Z' ) \times \mathcal{B}(Z'  \to \mu^+ \mu^-)$ \cite{LHCb:2016awg} where $Z' $ is a light ($0.2\,\lesssim m_{Z' }\,\lesssim 4 $ GeV) hidden sector boson produced as a resonance. Such experimental constraints are also relevant for our models of interest. The $U(1)_{B-3L_\mu}$ model, is especially compelling since it not only allows us to explain the observed anomalous magnetic moment of the muon\footnote{The same $\zbl$-mediated diagram (Fig.~\ref{fig:muong2}) will also contribute to $\Delta a_e$ in case of $U(1)_{B-3L_e}$ scenario, however, this contribution can not explain the relative sign in $\Delta a_e$.}, but could also have the potential to explain the anomalies in $b \to s \ell \ell$ decays\footnote{For studies of light vector boson models in the context of rare FCNC decays and $(g-2)_\mu$ see \cite{AristizabalSierra:2015vqb, Datta:2017pfz, Datta:2017ezo,  Sala:2017ihs, Kang:2019vng, Borah:2020swo, Cen:2021iwv}.}. However, the realm of the sub-GeV vector boson with $\gbl \lesssim 10^{-5}$, as required for the dark matter freeze-in, is beyond the reach of current experimental sensitivity. As already pointed out in~\cite{Bauer:2020itv}, future experimental searches would require a sensitivity of $\mathcal{B}(B \to K (e^+ e^- /\mu^+ \mu^-)) \leq 10^{-15}$ for $\mzbl \sim 1$ GeV in order to probe the relic favoured parameter space shown in Figs.~\ref{fig:relic1} and \ref{fig:relic2}. 

\section{Conclusion}
\label{sec:concl}
In this work we have explored the viable parameter space for freeze-in production of dark matter (DM) in flavoured gauge extensions of the SM, where the SM gauge symmetry is augmented by an abelian $U(1)_{B-3L_i}$ local gauge group, with $i\in e\,,\mu\,,\tau$. Apart from anomaly cancellation, three generations of right handed neutrinos (RHNs) take part in explaining the origin of small neutrino masses via Type-I seesaw. The DM is assumed to be a singlet vector-like neutral lepton that carries a non-trivial charge under the new gauge symmetry which in turn also protects the DM from decaying. The charges of the new particles are assigned in a way that apart from the heavy neutral gauge boson $\zbl$, the DM does not have any other portal to the visible sector. Thus, the DM phenomenology is decided by the new gauge coupling $\gbl$ and mass of the new gauge boson $\mzbl$, together with the DM charge $x$ and DM mass $m_\chi$.

Due to the universal coupling of $\zbl$, we employ the two-step freeze-in mechanism, where $\zbl$ is first produced from the thermal bath and then its further decay leads to the production of the DM $\chi$. Besides the decay, the particles in the bath can also undergo $2\to2$ annihilation mediated by $\zbl$, pair-producing the DM particles. We find, it is possible to satisfy the PLANCK observed relic density for $\mzbl\lesssim\mdm$ and $\mzbl\gtrsim 2\,\mdm$. Particularly, in the former scenario the DM abundance becomes independent of the mediator mass $m_{\zbl}$.    

For gauge coupling values so tiny as mandated from the requirement of the freeze-in criterion in the early Universe, existing experimental searches are not sensitive to flavor observables like, e.g., flavour changing transitions in the mass basis. However some constraints arise on flavour non-universal couplings through non-standard interactions (NSI) due to its couplings to baryons or electrons are present, as evident, e.g., from $U(1)_{B-3L_{\mu}}$ gauge boson from recent resonance searches for $B^+ \rightarrow K^+\mu^+\mu^-$ at the LHCb. We typically focus on the region of the parameter space where the mass $\mzbl\lesssim 10$ GeV, which is of interest in several energy and intensity frontier experiments. We find the DM parameter space becomes very tightly constrained for the case of $U(1)_{B-3L_e}$ and $U(1)_{B-3L_\mu}$ scenarios, while for $U(1)_{B-3L_\tau}$ the constraints are much relaxed as one can see from the bottom panel of Fig.~\ref{fig:relic1} and Fig.~\ref{fig:relic2}, depending on the size of $\gbl$. We show, experiments like FASER, FASER2, FASER$\nu$ and SHiP etc. will be able to probe a substantial part of the freeze-in parameter space for size of the BSM gauge coupling $\gbl\simeq\mathcal{O}(10^{-8})$ and for $\mzbl\sim 1$ GeV, while larger coupling values can be constrained from experiments like Belle-II, BaBar and from non-standard neutrino interactions (NSI). 

We thus conclude, in a well-motivated UV-complete model with possible explanation of the hierarchy of the flavor structure in the leptonic sector, one may envisage a flavorful vector-portal non-thermal DM candidate via freeze-in. Additionally, one will be able to probe the predicted tiny freeze-in couplings in several current and upcoming light dark world of experimental facilities, and see imprints of non-thermal DM which eluded detection prospects in the past with traditional DM direct searches. 
\section*{Acknowledgements}
BB received funding from the Patrimonio Autónomo - Fondo Nacional de Financiamiento para la Ciencia, la Tecnología y la Innovación Francisco José de Caldas (MinCiencias - Colombia) grant 80740-465-2020. This project has received funding /support from the European Union's Horizon 2020 research and innovation programme under the Marie Sklodowska-Curie grant agreement No 860881-HIDDeN. LM received funding from NSF Grant No. PHY1915142. PG would like to acknowledge the support from DAE, India for the Regional Centre for Accelerator based Particle Physics (RECAPP), Harish Chandra Research Institute. The authors would like to thank Sudip Jana, Soumitra Nandi and Nobuchika Okada for useful discussions, and Arindam Das and P S Bhupal Dev for providing valuable feedback on the manuscript.

\appendix
\section{$\zbl$ production from SM bath}
\label{sec:app-zbl-prod}
Here we have gathered the expressions for all decay and annihilation cross-sections contributing to $\zbl$ production, specifically after EWSB. The $h_{1,2}\to\zbl\zbl$ decay width reads

\begin{equation}
\Gamma_{h_{1(2)}\to\zbl\zbl}=\frac{9\,\gbl^2\,\sin^2(\cos^2)\theta}{32\,\pi\,r^2}\,\frac{\mzbl^2}{m_{1(2)}}\,\sqrt{1-4\,r^2}\,(1-4\,r^2+12\,r^4)\,;\,r=\mzbl/m_{1(2)}\,,    
\end{equation}

Next we have the most dominating annihilation channel $ff\to\zbl\,,\gamma$, $(f\in q,\ell)$, which give rise to a cross-section 

\begin{align}
&\sigma(s)_{qq\to\zbl\gamma}= \frac{\gbl^2\,m_W^4}{\mathcal{A}}\,s^{5/2}\,\sqrt{1-\frac{\mzbl^2}{s}}\,\Biggl(1-\frac{4\,m_q^2}{s}\Biggr)^{-1/4}\,\mathcal{B}\,,   
\end{align}

\noindent where

\begin{align}
& \mathcal{A} = 3888\,\pi^2\,m_Z^2\,\pi\,v_d^2\,\left(s-4 m_q^2\right)\,\left(s-\mzbl^2\right)^3\nonumber\\&\left[\mzbl^2 s-4 m_q^2 \left(\mzbl^2-s\right)\right]\,\left[4 m_q^2\,\left(\mzbl^2-s\right)+s \left(2 s-\mzbl^2\right)\right]\,,    
\end{align}

\noindent and

\begin{align}
&\mathcal{B}= \frac{\mathcal{C}}{s^2}+\frac{\mathcal{D}}{s^4}\,\log\Biggl[\frac{s \left(\mzbl^2-2 s\right)-4 m_q^2 \left(\mzbl^2-s\right)}{4 m_q^2 \left(\mzbl^2-s\right)-\mzbl^2 s}\Biggr]\,,   
\end{align}

\noindent with 

\begin{align}
&\mathcal{C}=\left(4 m_q^2-s\right) \left(s-\mzbl^2\right)^3\nonumber\\&\left[16 m_q^4-s^2 \left(\frac{\left(s-4 m_q^2\right)^2 \left(\mzbl^2-s\right)^4}{s^6}-1\right)+8 m_q^2 \mzbl^2+\mzbl^4-2 \mzbl^2 s\right]\,,  
\end{align}

\noindent and

\begin{align}
&\mathcal{D} = \left(-8 m_q^4-4 m_q^2 \left(\mzbl^2-s\right)+\mzbl^4+s^2\right) \Biggl[-16 m_q^4 \left(\mzbl^2-s\right)^4+8 m_q^2 s \left(\mzbl^2-s\right)^4\nonumber\\&-\mzbl^2 s^2 \left(\mzbl^2-2 s\right) \left(\mzbl^2-s\right)^2\Biggr]\,.    
\end{align}

The heavy gauge bosons of the SM can also give rise to  pair production of $\zbl$, which read 

\begin{align}
& \sigma(s)_{VV\to\zbl\zbl}\simeq\frac{\gbl^2\,m_W^4}{32\,\pi\,v_d^2\,s}\,\sqrt{\frac{s-4\,\mdm^2}{s-4\,m_V^2}}\,\frac{(s^2-4\,m_V^2\,s+12\,m_V^4)\,(s^2-4\,\mzbl^2\,s+12\,\mzbl^4)}{\mzbl^2\,(m_W^2-m_Z^2)^2}\nonumber\\&\Biggl[\frac{(m_1^2-m_2^2)}{(s-m_1^2)^2+\Gamma_1^2\,m_1^2}\Biggr]\,\Biggl[\frac{(m_1^2-m_2^2)}{(s-m_2^2)^2+\Gamma_2^2\,m_2^2}\Biggr]\,\sin^2\theta\,\cos^2\theta+\mathcal{O}[\gbl^4]\,, 
\end{align}

The contributions from the SM fermions give rise to 

\begin{align}
&\sigma(s)_{qq\to\zbl\zbl}\simeq\frac{3\,\gbl^2}{64\,\pi\,s}\,\cos^2\theta\,\sin^2\theta\,\frac{m_q^2\,m_W^2}{\mzbl^2\,v_d^2}\sqrt{(s-4\,\mzbl^2)(s-4\,m_q^2)}\nonumber\\&\Biggl[\frac{s^2-4\,s\,\mzbl^2+12\,\mzbl^4}{m_Z^2-m_W^2}\Biggr]\Biggl[\frac{(m_1^2-m_2^2)}{(s-m_1^2)^2+\Gamma_1^2\,m_1^2}\Biggr]\,\Biggl[\frac{(m_1^2-m_2^2)}{(s-m_2^2)^2+\Gamma_2^2\,m_2^2}\Biggr]+\mathcal{O}[\gbl^4]\,,    
\end{align}

\begin{align}
&\sigma(s)_{\ell\ell\to\zbl\zbl}\simeq\frac{9\,\gbl^2}{64\,\pi\,s}\,\cos^2\theta\,\sin^2\theta\,\frac{m_\ell^2\,m_W^2}{\mzbl^2\,v_d^2}\sqrt{(s-4\,\mzbl^2)(s-4\,m_q^2)}\nonumber\\&\Biggl[\frac{s^2-4\,s\,\mzbl^2+12\,\mzbl^4}{m_Z^2-m_W^2}\Biggr]\Biggl[\frac{(m_1^2-m_2^2)}{(s-m_1^2)^2+\Gamma_1^2\,m_1^2}\Biggr]\,\Biggl[\frac{(m_1^2-m_2^2)}{(s-m_2^2)^2+\Gamma_2^2\,m_2^2}\Biggr]+\mathcal{O}[\gbl^4]\,,    
\end{align}

Finally, the SM scalars can also contribute the $\zbl$ number density with cross-sections 

\begin{align}
&\sigma(s)_{h_{1(2)}h_{1(2)}\to\zbl\zbl}\simeq\frac{9\,s\,\gbl^2}{32\,\pi\,v_d^2}\,\Biggl(1+\frac{2\,m_{1(2)}^2}{s}\Biggr)^2\,\sqrt{\frac{s-4\,\mzbl^2}{s-4\,m_{1(2)}^2}}\,\sin^2(\cos^2)\theta\nonumber\\&\,\Biggl[\frac{s^2-4\,s\,\mzbl^2+12\,\mzbl^4}{\mzbl^2}\Biggr]\Biggl[\frac{(m_1^2-m_2^2)}{(s-m_1^2)^2+\Gamma_1^2\,m_1^2}\Biggr]\,\Biggl[\frac{(m_1^2-m_2^2)}{(s-m_2^2)^2+\Gamma_2^2\,m_2^2}\Biggr]+\mathcal{O}[\gbl^4]\,,    
\end{align}

\section{Decays and annhilations for freeze-in}
\label{sec:app-decay-ann}
The available decay channels for $\zbl$ are

\begin{align}
\Gamma_{\zbl\to qq} = \frac{\gbl^2\,\mzbl}{36\,\pi}\,\sqrt{1-4\,r^2}\,(1+2\,r^2)\,;\,r=m_q/\mzbl\,,    
\end{align}

\begin{align}
\Gamma_{\zbl\to\ell\ell} = \frac{3\,\gbl^2\,\mzbl}{4\,\pi}\,\sqrt{1-4\,r^2}\,(1+2\,r^2)\,;\,r=m_\ell/\mzbl\,,  \end{align}

\begin{equation}
\Gamma_{\zbl\to\chi\chi} = \frac{x^2\,\gbl^2\,\mzbl}{12\,\pi}\,\sqrt{1-4\,r^2}\,(1+2\,r^2)\,;r=\mdm/\mzbl\,.
\end{equation}

Annihilation cross-sections contributing to DM pair production read

\begin{align}
& \sigma(s)_{qq\to\chi\chi} = \frac{\gbl^4\,x^2\,s}{324\,\pi}\,\sqrt{\frac{s-4\,\mdm^2}{s-4\,m_q^2}}\,\Biggl[\frac{1}{(s-\mzbl^2)^2+\Gamma_{\zbl}^2\,\mzbl^2}\Biggr]\,\Biggl(1+\frac{2\,m_q^2}{s}\Biggr)\,\Biggl(1+\frac{2\,\mdm^2}{s}\Biggr)\,.   
\end{align}

\begin{align}
&\sigma(s)_{\ell\ell\to\chi\chi} =  \frac{3\,x^2\gbl^4\,s}{4\,\pi}\sqrt{\frac{s-4\,\mdm^2}{s-4\,m_\ell^2}}\,\Biggl[\frac{1}{(s-\mzbl^2)^2+\Gamma_{\zbl}^2\,\mzbl^2}\Biggr]\,\Biggl(1+\frac{2\,m_q^2}{s}\Biggr)\,\Biggl(1+\frac{2\,\mdm^2}{s}\Biggr)\,.   
\end{align}

\begin{align}
&\sigma(s)_{N_RN_R\to\chi\chi} =\frac{3\,x^2\,\gbl^4\,s}{8\,\pi}\,\sqrt{\frac{s-4\,\mdm^2}{s-4\,m_N^2}}\,\Biggl[\frac{1}{(s-\mzbl^2)^2+\Gamma_{\zbl}^2\,\mzbl^2}\Biggr]\,\Biggl(1-\frac{2\,m_N^2}{s}\Biggr)\,\Biggl(1+\frac{2\,\mdm^2}{s}\Biggr)\,. 
\end{align}

\section{Muon anomalous magnetic moment}
\label{sec:muong2}
\begin{figure}[htb!]
\centering
\includegraphics[scale=0.75]{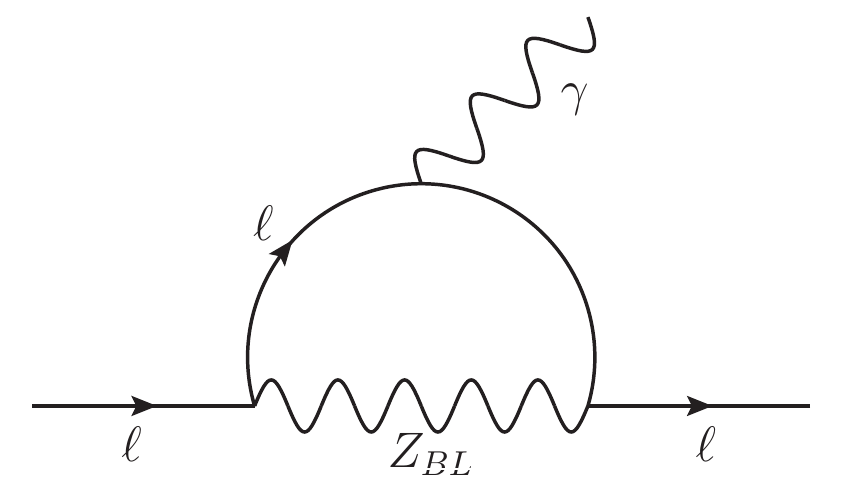}
\caption{$\zbl$ mediated one loop contribution to $\Delta a_\mu$.}
\label{fig:muong2}
\end{figure}

The contribution to the anomalous magnetic moment of the muon from the one loop diagram shown in Fig.~\ref{fig:muong2} is given by

\begin{equation}
\Delta a_\mu = \frac{9\,\gbl^2\, m_\mu^2}{8\,\pi^2}\,\int_0^1 dx \frac{2\,x^2\, (1-x)}{(1-x)\,\mzbl^2+ x^2\, m_\mu^2}\,.
\label{g2_Zprime}
\end{equation}

\bibliographystyle{JHEP}
\bibliography{Bibliography}

\providecommand{\href}[2]{#2}\begingroup\raggedright\begin{thebibliography}{100}

\bibitem{Khalil:2008kp}
S.~Khalil and O.~Seto, \emph{{Sterile neutrino dark matter in B - L extension
  of the standard model and galactic 511-keV line}},
  \href{https://doi.org/10.1088/1475-7516/2008/10/024}{\emph{JCAP} {\bfseries
  10} (2008) 024} [\href{https://arxiv.org/abs/0804.0336}{{\ttfamily
  0804.0336}}].

\bibitem{Okada:2010wd}
N.~Okada and O.~Seto, \emph{{Higgs portal dark matter in the minimal gauged
  $U(1)_{B-L}$ model}},
  \href{https://doi.org/10.1103/PhysRevD.82.023507}{\emph{Phys. Rev. D}
  {\bfseries 82} (2010) 023507}
  [\href{https://arxiv.org/abs/1002.2525}{{\ttfamily 1002.2525}}].

\bibitem{Kanemura:2011vm}
S.~Kanemura, O.~Seto and T.~Shimomura, \emph{{Masses of dark matter and
  neutrino from TeV scale spontaneous $U(1)_{B-L}$ breaking}},
  \href{https://doi.org/10.1103/PhysRevD.84.016004}{\emph{Phys. Rev. D}
  {\bfseries 84} (2011) 016004}
  [\href{https://arxiv.org/abs/1101.5713}{{\ttfamily 1101.5713}}].

\bibitem{Lindner:2011it}
M.~Lindner, D.~Schmidt and T.~Schwetz, \emph{{Dark Matter and neutrino masses
  from global U(1)$_{B−L}$ symmetry breaking}},
  \href{https://doi.org/10.1016/j.physletb.2011.10.022}{\emph{Phys. Lett. B}
  {\bfseries 705} (2011) 324}
  [\href{https://arxiv.org/abs/1105.4626}{{\ttfamily 1105.4626}}].

\bibitem{Okada:2012sg}
N.~Okada and Y.~Orikasa, \emph{{Dark matter in the classically conformal B-L
  model}}, \href{https://doi.org/10.1103/PhysRevD.85.115006}{\emph{Phys. Rev.
  D} {\bfseries 85} (2012) 115006}
  [\href{https://arxiv.org/abs/1202.1405}{{\ttfamily 1202.1405}}].

\bibitem{Basak:2013cga}
T.~Basak and T.~Mondal, \emph{{Constraining Minimal $U(1)_{B-L}$ model from
  Dark Matter Observations}},
  \href{https://doi.org/10.1103/PhysRevD.89.063527}{\emph{Phys. Rev. D}
  {\bfseries 89} (2014) 063527}
  [\href{https://arxiv.org/abs/1308.0023}{{\ttfamily 1308.0023}}].

\bibitem{Kanemura:2014rpa}
S.~Kanemura, T.~Matsui and H.~Sugiyama, \emph{{Neutrino mass and dark matter
  from gauged $U(1)_{B-L}$ breaking}},
  \href{https://doi.org/10.1103/PhysRevD.90.013001}{\emph{Phys. Rev. D}
  {\bfseries 90} (2014) 013001}
  [\href{https://arxiv.org/abs/1405.1935}{{\ttfamily 1405.1935}}].

\bibitem{Okada:2016gsh}
N.~Okada and S.~Okada, \emph{{$Z^\prime_{BL}$ portal dark matter and LHC Run-2
  results}}, \href{https://doi.org/10.1103/PhysRevD.93.075003}{\emph{Phys. Rev.
  D} {\bfseries 93} (2016) 075003}
  [\href{https://arxiv.org/abs/1601.07526}{{\ttfamily 1601.07526}}].

\bibitem{Biswas:2016ewm}
A.~Biswas, S.~Choubey and S.~Khan, \emph{{Galactic gamma ray excess and dark
  matter phenomenology in a $U(1)_{B-L}$ model}},
  \href{https://doi.org/10.1007/JHEP08(2016)114}{\emph{JHEP} {\bfseries 08}
  (2016) 114} [\href{https://arxiv.org/abs/1604.06566}{{\ttfamily
  1604.06566}}].

\bibitem{Nanda:2017bmi}
D.~Nanda and D.~Borah, \emph{{Common origin of neutrino mass and dark matter
  from anomaly cancellation requirements of a $U(1)_{B-L}$ model}},
  \href{https://doi.org/10.1103/PhysRevD.96.115014}{\emph{Phys. Rev. D}
  {\bfseries 96} (2017) 115014}
  [\href{https://arxiv.org/abs/1709.08417}{{\ttfamily 1709.08417}}].

\bibitem{Singirala:2017cch}
S.~Singirala, R.~Mohanta, S.~Patra and S.~Rao, \emph{{Majorana Dark Matter in a
  new $B−L$ model}},
  \href{https://doi.org/10.1088/1475-7516/2018/11/026}{\emph{JCAP} {\bfseries
  11} (2018) 026} [\href{https://arxiv.org/abs/1710.05775}{{\ttfamily
  1710.05775}}].

\bibitem{Bandyopadhyay:2017bgh}
P.~Bandyopadhyay, E.J.~Chun and R.~Mandal, \emph{{Implications of right-handed
  neutrinos in $B-L$ extended standard model with scalar dark matter}},
  \href{https://doi.org/10.1103/PhysRevD.97.015001}{\emph{Phys. Rev. D}
  {\bfseries 97} (2018) 015001}
  [\href{https://arxiv.org/abs/1707.00874}{{\ttfamily 1707.00874}}].

\bibitem{DeRomeri:2017oxa}
V.~De~Romeri, E.~Fernandez-Martinez, J.~Gehrlein, P.A.N.~Machado and V.~Niro,
  \emph{{Dark Matter and the elusive $Z′$ in a dynamical Inverse Seesaw
  scenario}}, \href{https://doi.org/10.1007/JHEP10(2017)169}{\emph{JHEP}
  {\bfseries 10} (2017) 169}
  [\href{https://arxiv.org/abs/1707.08606}{{\ttfamily 1707.08606}}].

\bibitem{Borah:2018smz}
D.~Borah, D.~Nanda, N.~Narendra and N.~Sahu, \emph{{Right-handed neutrino dark
  matter with radiative neutrino mass in gauged B \ensuremath{-} L model}},
  \href{https://doi.org/10.1016/j.nuclphysb.2019.114841}{\emph{Nucl. Phys. B}
  {\bfseries 950} (2020) 114841}
  [\href{https://arxiv.org/abs/1810.12920}{{\ttfamily 1810.12920}}].

\bibitem{Okada:2018tgy}
N.~Okada, S.~Okada and D.~Raut, \emph{{Natural Z' -portal Majorana dark matter
  in alternative U(1) extended standard model}},
  \href{https://doi.org/10.1103/PhysRevD.100.035022}{\emph{Phys. Rev. D}
  {\bfseries 100} (2019) 035022}
  [\href{https://arxiv.org/abs/1811.11927}{{\ttfamily 1811.11927}}].

\bibitem{Das:2018tbd}
A.~Das, N.~Okada, S.~Okada and D.~Raut, \emph{{Probing the seesaw mechanism at
  the 250 GeV ILC}},
  \href{https://doi.org/10.1016/j.physletb.2019.134849}{\emph{Phys. Lett. B}
  {\bfseries 797} (2019) 134849}
  [\href{https://arxiv.org/abs/1812.11931}{{\ttfamily 1812.11931}}].

\bibitem{Biswas:2019ygr}
A.~Biswas, D.~Borah and D.~Nanda, \emph{{Type III seesaw for neutrino masses in
  U(1)$_{B−L}$ model with multi-component dark matter}},
  \href{https://doi.org/10.1007/JHEP12(2019)109}{\emph{JHEP} {\bfseries 12}
  (2019) 109} [\href{https://arxiv.org/abs/1908.04308}{{\ttfamily
  1908.04308}}].

\bibitem{Das:2019pua}
A.~Das, S.~Goswami, K.N.~Vishnudath and T.~Nomura, \emph{{Constraining a
  general U(1)$^\prime$ inverse seesaw model from vacuum stability, dark matter
  and collider}},
  \href{https://doi.org/10.1103/PhysRevD.101.055026}{\emph{Phys. Rev. D}
  {\bfseries 101} (2020) 055026}
  [\href{https://arxiv.org/abs/1905.00201}{{\ttfamily 1905.00201}}].

\bibitem{Baules:2019zwk}
V.~Baules, N.~Okada and S.~Okada, \emph{{Braneworld Cosmological Effect on
  Freeze-in Dark Matter Density and Lifetime Frontier}},
  \href{https://arxiv.org/abs/1911.05344}{{\ttfamily 1911.05344}}.

\bibitem{Mohapatra:2019ysk}
R.N.~Mohapatra and N.~Okada, \emph{{Dark Matter Constraints on Low Mass and
  Weakly Coupled B-L Gauge Boson}},
  \href{https://doi.org/10.1103/PhysRevD.102.035028}{\emph{Phys. Rev. D}
  {\bfseries 102} (2020) 035028}
  [\href{https://arxiv.org/abs/1908.11325}{{\ttfamily 1908.11325}}].

\bibitem{Mohapatra:2020bze}
R.N.~Mohapatra and N.~Okada, \emph{{Freeze-in Dark Matter from a Minimal B-L
  Model and Possible Grand Unification}},
  \href{https://doi.org/10.1103/PhysRevD.101.115022}{\emph{Phys. Rev. D}
  {\bfseries 101} (2020) 115022}
  [\href{https://arxiv.org/abs/2005.00365}{{\ttfamily 2005.00365}}].

\bibitem{Dutta:2020xwn}
M.~Dutta, S.~Bhattacharya, P.~Ghosh and N.~Sahu, \emph{{Singlet-Doublet
  Majorana Dark Matter and Neutrino Mass in a minimal Type-I Seesaw Scenario}},
  \href{https://doi.org/10.1088/1475-7516/2021/03/008}{\emph{JCAP} {\bfseries
  03} (2021) 008} [\href{https://arxiv.org/abs/2009.00885}{{\ttfamily
  2009.00885}}].

\bibitem{Biswas:2020ldp}
A.~Biswas, D.~Borah and D.~Nanda, \emph{{Type III Seesaw and Two-Component Dark
  Matter in $U(1)_{B-L}$ Model}},
  \href{https://doi.org/10.1007/978-981-15-6292-1_36}{\emph{Springer Proc.
  Phys.} {\bfseries 248} (2020) 297}.

\bibitem{Okada:2020evk}
N.~Okada, S.~Okada, D.~Raut and Q.~Shafi, \emph{{Dark matter $Z^\prime$ and
  XENON1T excess from $U(1)_X$ extended standard model}},
  \href{https://doi.org/10.1016/j.physletb.2020.135785}{\emph{Phys. Lett. B}
  {\bfseries 810} (2020) 135785}
  [\href{https://arxiv.org/abs/2007.02898}{{\ttfamily 2007.02898}}].

\bibitem{Ghosh:2021khk}
P.~Ghosh, S.~Mahapatra, N.~Narendra and N.~Sahu, \emph{{Experimentally
  Verifiable $U(1)_{ B-L}$ Symmetric Model with Type-II Seesaw and Dark
  Matter}},  \href{https://arxiv.org/abs/2107.11951}{{\ttfamily 2107.11951}}.

\bibitem{Nath:2021uqb}
N.~Nath, N.~Okada, S.~Okada, D.~Raut and Q.~Shafi, \emph{{Light $Z^\prime$ and
  Dirac fermion dark matter in the $B-L$ model}},
  \href{https://arxiv.org/abs/2112.08960}{{\ttfamily 2112.08960}}.

\bibitem{He:1991qd}
X.-G.~He, G.C.~Joshi, H.~Lew and R.R.~Volkas, \emph{{Simplest Z-prime model}},
  \href{https://doi.org/10.1103/PhysRevD.44.2118}{\emph{Phys. Rev. D}
  {\bfseries 44} (1991) 2118}.

\bibitem{PhysRevD.64.055006}
S.~Baek, N.G.~Deshpande, X.-G.~He and P.~Ko, \emph{Muon anomalous
  $g\ensuremath{-}2$ and gauged
  ${L}_{\ensuremath{\mu}}{\ensuremath{-}l}_{\ensuremath{\tau}}$ models},
  \href{https://doi.org/10.1103/PhysRevD.64.055006}{\emph{Phys. Rev. D}
  {\bfseries 64} (2001) 055006}.

\bibitem{Ma:2001md}
E.~Ma, D.P.~Roy and S.~Roy, \emph{{Gauged L(mu) - L(tau) with large muon
  anomalous magnetic moment and the bimaximal mixing of neutrinos}},
  \href{https://doi.org/10.1016/S0370-2693(01)01428-9}{\emph{Phys. Lett. B}
  {\bfseries 525} (2002) 101}
  [\href{https://arxiv.org/abs/hep-ph/0110146}{{\ttfamily hep-ph/0110146}}].

\bibitem{Salvioni:2009jp}
E.~Salvioni, A.~Strumia, G.~Villadoro and F.~Zwirner, \emph{{Non-universal
  minimal Z' models: present bounds and early LHC reach}},
  \href{https://doi.org/10.1007/JHEP03(2010)010}{\emph{JHEP} {\bfseries 03}
  (2010) 010} [\href{https://arxiv.org/abs/0911.1450}{{\ttfamily 0911.1450}}].

\bibitem{Heeck:2011wj}
J.~Heeck and W.~Rodejohann, \emph{{Gauged $L_\mu - L_\tau$ Symmetry at the
  Electroweak Scale}},
  \href{https://doi.org/10.1103/PhysRevD.84.075007}{\emph{Phys. Rev. D}
  {\bfseries 84} (2011) 075007}
  [\href{https://arxiv.org/abs/1107.5238}{{\ttfamily 1107.5238}}].

\bibitem{Harigaya:2013twa}
K.~Harigaya, T.~Igari, M.M.~Nojiri, M.~Takeuchi and K.~Tobe, \emph{{Muon g-2
  and LHC phenomenology in the $L_\mu-L_\tau$ gauge symmetric model}},
  \href{https://doi.org/10.1007/JHEP03(2014)105}{\emph{JHEP} {\bfseries 03}
  (2014) 105} [\href{https://arxiv.org/abs/1311.0870}{{\ttfamily 1311.0870}}].

\bibitem{CARONE2013118}
C.D.~Carone, \emph{Flavor-nonuniversal dark gauge bosons and the muon g−2},
  \href{https://doi.org/https://doi.org/10.1016/j.physletb.2013.03.011}{\emph{Physics
  Letters B} {\bfseries 721} (2013) 118}.

\bibitem{Altmannshofer:2014cfa}
W.~Altmannshofer, S.~Gori, M.~Pospelov and I.~Yavin, \emph{{Quark flavor
  transitions in $L_\mu-L_\tau$ models}},
  \href{https://doi.org/10.1103/PhysRevD.89.095033}{\emph{Phys. Rev. D}
  {\bfseries 89} (2014) 095033}
  [\href{https://arxiv.org/abs/1403.1269}{{\ttfamily 1403.1269}}].

\bibitem{Farzan:2015doa}
Y.~Farzan, \emph{{A model for large non-standard interactions of neutrinos
  leading to the LMA-Dark solution}},
  \href{https://doi.org/10.1016/j.physletb.2015.07.015}{\emph{Phys. Lett. B}
  {\bfseries 748} (2015) 311}
  [\href{https://arxiv.org/abs/1505.06906}{{\ttfamily 1505.06906}}].

\bibitem{Farzan:2015hkd}
Y.~Farzan and I.M.~Shoemaker, \emph{{Lepton Flavor Violating Non-Standard
  Interactions via Light Mediators}},
  \href{https://doi.org/10.1007/JHEP07(2016)033}{\emph{JHEP} {\bfseries 07}
  (2016) 033} [\href{https://arxiv.org/abs/1512.09147}{{\ttfamily
  1512.09147}}].

\bibitem{Biswas:2016yjr}
A.~Biswas, S.~Choubey and S.~Khan, \emph{{FIMP and Muon ($g-2$) in a
  U$(1)_{L_{\mu}-L_{\tau}}$ Model}},
  \href{https://doi.org/10.1007/JHEP02(2017)123}{\emph{JHEP} {\bfseries 02}
  (2017) 123} [\href{https://arxiv.org/abs/1612.03067}{{\ttfamily
  1612.03067}}].

\bibitem{Biswas:2019twf}
A.~Biswas and A.~Shaw, \emph{{Reconciling dark matter, $R_{K^{(*)}}$ anomalies
  and $(g-2)_{\mu}$ in an ${L_{\mu}-L_{\tau}}$ scenario}},
  \href{https://doi.org/10.1007/JHEP05(2019)165}{\emph{JHEP} {\bfseries 05}
  (2019) 165} [\href{https://arxiv.org/abs/1903.08745}{{\ttfamily
  1903.08745}}].

\bibitem{Bauer:2020itv}
M.~Bauer, P.~Foldenauer and M.~Mosny, \emph{{Flavor structure of anomaly-free
  hidden photon models}},
  \href{https://doi.org/10.1103/PhysRevD.103.075024}{\emph{Phys. Rev. D}
  {\bfseries 103} (2021) 075024}
  [\href{https://arxiv.org/abs/2011.12973}{{\ttfamily 2011.12973}}].

\bibitem{Ma:1997nq}
E.~Ma, \emph{{Gauged B - 3L(tau) and radiative neutrino masses}},
  \href{https://doi.org/10.1016/S0370-2693(98)00599-1}{\emph{Phys. Lett. B}
  {\bfseries 433} (1998) 74}
  [\href{https://arxiv.org/abs/hep-ph/9709474}{{\ttfamily hep-ph/9709474}}].

\bibitem{Ma:1998dr}
E.~Ma and U.~Sarkar, \emph{{Gauged B - 3L(tau) and baryogenesis}},
  \href{https://doi.org/10.1016/S0370-2693(98)01019-3}{\emph{Phys. Lett. B}
  {\bfseries 439} (1998) 95}
  [\href{https://arxiv.org/abs/hep-ph/9807307}{{\ttfamily hep-ph/9807307}}].

\bibitem{Chang:2000xy}
L.N.~Chang, O.~Lebedev, W.~Loinaz and T.~Takeuchi, \emph{{Constraints on gauged
  B - 3 L(tau) and related theories}},
  \href{https://doi.org/10.1103/PhysRevD.63.074013}{\emph{Phys. Rev. D}
  {\bfseries 63} (2001) 074013}
  [\href{https://arxiv.org/abs/hep-ph/0010118}{{\ttfamily hep-ph/0010118}}].

\bibitem{Pal:2003ip}
P.B.~Pal and U.~Sarkar, \emph{{Gauged B - 3L(tau), low-energy unification and
  proton decay}},
  \href{https://doi.org/10.1016/j.physletb.2003.08.034}{\emph{Phys. Lett. B}
  {\bfseries 573} (2003) 147}
  [\href{https://arxiv.org/abs/hep-ph/0306088}{{\ttfamily hep-ph/0306088}}].

\bibitem{Chang:2009tx}
Q.~Chang, X.-Q.~Li and Y.-D.~Yang, \emph{{Family Non-universal Z-prime effects
  on anti-B(q) - B(q) mixing, B ---\ensuremath{>} X(s) mu+ mu- and B(s)
  ---\ensuremath{>} mu+ mu- Decays}},
  \href{https://doi.org/10.1007/JHEP02(2010)082}{\emph{JHEP} {\bfseries 02}
  (2010) 082} [\href{https://arxiv.org/abs/0907.4408}{{\ttfamily 0907.4408}}].

\bibitem{Lee:2010hf}
H.-S.~Lee and E.~Ma, \emph{{Gauged $B-x_i L$ origin of $R$ Parity and its
  implications}},
  \href{https://doi.org/10.1016/j.physletb.2010.04.032}{\emph{Phys. Lett. B}
  {\bfseries 688} (2010) 319}
  [\href{https://arxiv.org/abs/1001.0768}{{\ttfamily 1001.0768}}].

\bibitem{Okada:2012sp}
H.~Okada, \emph{{Dark Matters in Gauged $B-3L_i$ Model}},
  \href{https://arxiv.org/abs/1212.0492}{{\ttfamily 1212.0492}}.

\bibitem{Chun:2018ibr}
E.J.~Chun, A.~Das, J.~Kim and J.~Kim, \emph{{Searching for flavored gauge
  bosons}}, \href{https://doi.org/10.1007/JHEP02(2019)093}{\emph{JHEP}
  {\bfseries 02} (2019) 093}
  [\href{https://arxiv.org/abs/1811.04320}{{\ttfamily 1811.04320}}].

\bibitem{Allanach:2018lvl}
B.C.~Allanach and J.~Davighi, \emph{{Third family hypercharge model for $
  {R}_{K^{\left(\ast \right)}} $ and aspects of the fermion mass problem}},
  \href{https://doi.org/10.1007/JHEP12(2018)075}{\emph{JHEP} {\bfseries 12}
  (2018) 075} [\href{https://arxiv.org/abs/1809.01158}{{\ttfamily
  1809.01158}}].

\bibitem{Borah:2021jzu}
D.~Borah, M.~Dutta, S.~Mahapatra and N.~Sahu, \emph{{Muon (g-2) and XENON1T
  excess with boosted dark matter in $U(1)_{L_{\mu}-L_{\tau}}$ model}},
  \href{https://doi.org/10.1016/j.physletb.2021.136577}{\emph{Phys. Lett. B}
  {\bfseries 820} (2021) 136577}
  [\href{https://arxiv.org/abs/2104.05656}{{\ttfamily 2104.05656}}].

\bibitem{Borah:2021khc}
D.~Borah, M.~Dutta, S.~Mahapatra and N.~Sahu, \emph{{Lepton Anomalous Magnetic
  Moment with Singlet-Doublet Fermion Dark Matter in Scotogenic
  $U(1)_{L_{\mu}-L_{\tau}}$ Model}},
  \href{https://arxiv.org/abs/2109.02699}{{\ttfamily 2109.02699}}.

\bibitem{Zwicky:1933gu}
F.~Zwicky, \emph{{Die Rotverschiebung von extragalaktischen Nebeln}},
  \href{https://doi.org/10.1007/s10714-008-0707-4}{\emph{Helv. Phys. Acta}
  {\bfseries 6} (1933) 110}.

\bibitem{Zwicky:1937zza}
F.~Zwicky, \emph{{On the Masses of Nebulae and of Clusters of Nebulae}},
  \href{https://doi.org/10.1086/143864}{\emph{Astrophys. J.} {\bfseries 86}
  (1937) 217}.

\bibitem{Rubin:1970zza}
V.C.~Rubin and W.K.~Ford, Jr., \emph{{Rotation of the Andromeda Nebula from a
  Spectroscopic Survey of Emission Regions}},
  \href{https://doi.org/10.1086/150317}{\emph{Astrophys. J.} {\bfseries 159}
  (1970) 379}.

\bibitem{Clowe:2006eq}
D.~Clowe, M.~Brada\v{c}, A.H.~Gonzalez, M.~Markevitch, S.W.~Randall, C.~Jones
  et~al., \emph{{A direct empirical proof of the existence of dark matter}},
  \href{https://doi.org/10.1086/508162}{\emph{Astrophys. J. Lett.} {\bfseries
  648} (2006) L109} [\href{https://arxiv.org/abs/astro-ph/0608407}{{\ttfamily
  astro-ph/0608407}}].

\bibitem{Hu:2001bc}
W.~Hu and S.~Dodelson, \emph{{Cosmic microwave background anisotropies}},
  \href{https://doi.org/10.1146/annurev.astro.40.060401.093926}{\emph{Ann. Rev.
  Astron. Astrophys.} {\bfseries 40} (2002) 171}
  [\href{https://arxiv.org/abs/astro-ph/0110414}{{\ttfamily
  astro-ph/0110414}}].

\bibitem{Aghanim:2018eyx}
{\scshape Planck} collaboration, \emph{{Planck 2018 results. VI. Cosmological
  parameters}},
  \href{https://doi.org/10.1051/0004-6361/201833910}{\emph{Astron. Astrophys.}
  {\bfseries 641} (2020) A6}
  [\href{https://arxiv.org/abs/1807.06209}{{\ttfamily 1807.06209}}].

\bibitem{Jungman:1995df}
G.~Jungman, M.~Kamionkowski and K.~Griest, \emph{{Supersymmetric dark matter}},
  \href{https://doi.org/10.1016/0370-1573(95)00058-5}{\emph{Phys. Rept.}
  {\bfseries 267} (1996) 195}
  [\href{https://arxiv.org/abs/hep-ph/9506380}{{\ttfamily hep-ph/9506380}}].

\bibitem{Bertone:2004pz}
G.~Bertone, D.~Hooper and J.~Silk, \emph{{Particle dark matter: Evidence,
  candidates and constraints}},
  \href{https://doi.org/10.1016/j.physrep.2004.08.031}{\emph{Phys. Rept.}
  {\bfseries 405} (2005) 279}
  [\href{https://arxiv.org/abs/hep-ph/0404175}{{\ttfamily hep-ph/0404175}}].

\bibitem{Feng:2010gw}
J.L.~Feng, \emph{{Dark Matter Candidates from Particle Physics and Methods of
  Detection}},
  \href{https://doi.org/10.1146/annurev-astro-082708-101659}{\emph{Ann. Rev.
  Astron. Astrophys.} {\bfseries 48} (2010) 495}
  [\href{https://arxiv.org/abs/1003.0904}{{\ttfamily 1003.0904}}].

\bibitem{XENON:2018voc}
{\scshape XENON} collaboration, \emph{{Dark Matter Search Results from a One
  Ton-Year Exposure of XENON1T}},
  \href{https://doi.org/10.1103/PhysRevLett.121.111302}{\emph{Phys. Rev. Lett.}
  {\bfseries 121} (2018) 111302}
  [\href{https://arxiv.org/abs/1805.12562}{{\ttfamily 1805.12562}}].

\bibitem{XENON:2020kmp}
{\scshape XENON} collaboration, \emph{{Projected WIMP sensitivity of the
  XENONnT dark matter experiment}},
  \href{https://doi.org/10.1088/1475-7516/2020/11/031}{\emph{JCAP} {\bfseries
  11} (2020) 031} [\href{https://arxiv.org/abs/2007.08796}{{\ttfamily
  2007.08796}}].

\bibitem{PandaX:2018wtu}
{\scshape PandaX} collaboration, \emph{{Dark matter direct search sensitivity
  of the PandaX-4T experiment}},
  \href{https://doi.org/10.1007/s11433-018-9259-0}{\emph{Sci. China Phys. Mech.
  Astron.} {\bfseries 62} (2019) 31011}
  [\href{https://arxiv.org/abs/1806.02229}{{\ttfamily 1806.02229}}].

\bibitem{McDonald:2001vt}
J.~McDonald, \emph{{Thermally generated gauge singlet scalars as
  selfinteracting dark matter}},
  \href{https://doi.org/10.1103/PhysRevLett.88.091304}{\emph{Phys. Rev. Lett.}
  {\bfseries 88} (2002) 091304}
  [\href{https://arxiv.org/abs/hep-ph/0106249}{{\ttfamily hep-ph/0106249}}].

\bibitem{Hall:2009bx}
L.J.~Hall, K.~Jedamzik, J.~March-Russell and S.M.~West, \emph{{Freeze-In
  Production of FIMP Dark Matter}},
  \href{https://doi.org/10.1007/JHEP03(2010)080}{\emph{JHEP} {\bfseries 03}
  (2010) 080} [\href{https://arxiv.org/abs/0911.1120}{{\ttfamily 0911.1120}}].

\bibitem{Chu:2011be}
X.~Chu, T.~Hambye and M.H.G.~Tytgat, \emph{{The Four Basic Ways of Creating
  Dark Matter Through a Portal}},
  \href{https://doi.org/10.1088/1475-7516/2012/05/034}{\emph{JCAP} {\bfseries
  05} (2012) 034} [\href{https://arxiv.org/abs/1112.0493}{{\ttfamily
  1112.0493}}].

\bibitem{Bernal:2017kxu}
N.~Bernal, M.~Heikinheimo, T.~Tenkanen, K.~Tuominen and V.~Vaskonen, \emph{{The
  Dawn of FIMP Dark Matter: A Review of Models and Constraints}},
  \href{https://doi.org/10.1142/S0217751X1730023X}{\emph{Int. J. Mod. Phys. A}
  {\bfseries 32} (2017) 1730023}
  [\href{https://arxiv.org/abs/1706.07442}{{\ttfamily 1706.07442}}].

\bibitem{Duch:2017khv}
M.~Duch, B.~Grzadkowski and D.~Huang, \emph{{Strongly self-interacting vector
  dark matter via freeze-in}},
  \href{https://doi.org/10.1007/JHEP01(2018)020}{\emph{JHEP} {\bfseries 01}
  (2018) 020} [\href{https://arxiv.org/abs/1710.00320}{{\ttfamily
  1710.00320}}].

\bibitem{Biswas:2018aib}
A.~Biswas, D.~Borah and A.~Dasgupta, \emph{{UV complete framework of freeze-in
  massive particle dark matter}},
  \href{https://doi.org/10.1103/PhysRevD.99.015033}{\emph{Phys. Rev. D}
  {\bfseries 99} (2019) 015033}
  [\href{https://arxiv.org/abs/1805.06903}{{\ttfamily 1805.06903}}].

\bibitem{Heeba:2018wtf}
S.~Heeba, F.~Kahlhoefer and P.~St\"ocker, \emph{{Freeze-in production of
  decaying dark matter in five steps}},
  \href{https://doi.org/10.1088/1475-7516/2018/11/048}{\emph{JCAP} {\bfseries
  11} (2018) 048} [\href{https://arxiv.org/abs/1809.04849}{{\ttfamily
  1809.04849}}].

\bibitem{Barman:2019lvm}
B.~Barman, S.~Bhattacharya and M.~Zakeri, \emph{{Non-Abelian Vector Boson as
  FIMP Dark Matter}},
  \href{https://doi.org/10.1088/1475-7516/2020/02/029}{\emph{JCAP} {\bfseries
  02} (2020) 029} [\href{https://arxiv.org/abs/1905.07236}{{\ttfamily
  1905.07236}}].

\bibitem{Barman:2021lot}
B.~Barman and A.~Ghoshal, \emph{{Scale Invariant FIMP Miracle}},
  \href{https://arxiv.org/abs/2109.03259}{{\ttfamily 2109.03259}}.

\bibitem{Barman:2022njh}
B.~Barman and A.~Ghoshal, \emph{{Probing pre-BBN era with Scale Invarint
  FIMP}},  \href{https://arxiv.org/abs/2203.13269}{{\ttfamily 2203.13269}}.

\bibitem{Burell:2011wh}
Z.M.~Burell and N.~Okada, \emph{{Supersymmetric minimal B-L model at the TeV
  scale with right-handed Majorana neutrino dark matter}},
  \href{https://doi.org/10.1103/PhysRevD.85.055011}{\emph{Phys. Rev. D}
  {\bfseries 85} (2012) 055011}
  [\href{https://arxiv.org/abs/1111.1789}{{\ttfamily 1111.1789}}].

\bibitem{Okada:2020cue}
N.~Okada, S.~Okada and Q.~Shafi, \emph{{Light $Z′$ and dark matter from
  U(1)$_X$ gauge symmetry}},
  \href{https://doi.org/10.1016/j.physletb.2020.135845}{\emph{Phys. Lett. B}
  {\bfseries 810} (2020) 135845}
  [\href{https://arxiv.org/abs/2003.02667}{{\ttfamily 2003.02667}}].

\bibitem{delaVega:2021wpx}
L.M.G.~de~la Vega, L.J.~Flores, N.~Nath and E.~Peinado, \emph{{Complementarity
  between dark matter direct searches and CE\ensuremath{\nu}NS experiments in
  U(1)' models}}, \href{https://doi.org/10.1007/JHEP09(2021)146}{\emph{JHEP}
  {\bfseries 09} (2021) 146}
  [\href{https://arxiv.org/abs/2107.04037}{{\ttfamily 2107.04037}}].

\bibitem{Ghoshal:2022zwu}
A.~Ghoshal, N.~Okada and A.~Paul, \emph{{eV Hubble Scale Inflation with
  Radiative Plateau: Very light Inflaton, Reheating \& Dark Matter in $B-L$
  Extensions}},  \href{https://arxiv.org/abs/2203.03670}{{\ttfamily
  2203.03670}}.

\bibitem{Paul:2018njm}
A.~Paul, A.~Ghoshal, A.~Chatterjee and S.~Pal, \emph{{Inflation, (P)reheating
  and Neutrino Anomalies: Production of Sterile Neutrinos with Secret
  Interactions}},
  \href{https://doi.org/10.1140/epjc/s10052-019-7348-5}{\emph{Eur. Phys. J. C}
  {\bfseries 79} (2019) 818}
  [\href{https://arxiv.org/abs/1808.09706}{{\ttfamily 1808.09706}}].

\bibitem{Paul:2021ewd}
A.~Paul, A.~Chatterjee, A.~Ghoshal and S.~Pal, \emph{{Shedding Light on Dark
  Matter and Neutrino Interactions from Cosmology}},
  \href{https://arxiv.org/abs/2104.04760}{{\ttfamily 2104.04760}}.

\bibitem{Okada:2018ktp}
S.~Okada, \emph{{$Z'$ Portal Dark Matter in the Minimal $B-L$ Model}},
  \href{https://doi.org/10.1155/2018/5340935}{\emph{Adv. High Energy Phys.}
  {\bfseries 2018} (2018) 5340935}
  [\href{https://arxiv.org/abs/1803.06793}{{\ttfamily 1803.06793}}].

\bibitem{Robens:2015gla}
T.~Robens and T.~Stefaniak, \emph{{Status of the Higgs Singlet Extension of the
  Standard Model after LHC Run 1}},
  \href{https://doi.org/10.1140/epjc/s10052-015-3323-y}{\emph{Eur. Phys. J. C}
  {\bfseries 75} (2015) 104}
  [\href{https://arxiv.org/abs/1501.02234}{{\ttfamily 1501.02234}}].

\bibitem{Robens:2016xkb}
T.~Robens and T.~Stefaniak, \emph{{LHC Benchmark Scenarios for the Real Higgs
  Singlet Extension of the Standard Model}},
  \href{https://doi.org/10.1140/epjc/s10052-016-4115-8}{\emph{Eur. Phys. J. C}
  {\bfseries 76} (2016) 268}
  [\href{https://arxiv.org/abs/1601.07880}{{\ttfamily 1601.07880}}].

\bibitem{Chalons:2016jeu}
G.~Chalons, D.~Lopez-Val, T.~Robens and T.~Stefaniak, \emph{{The Higgs singlet
  extension at LHC Run 2}},
  \href{https://doi.org/10.22323/1.282.1180}{\emph{PoS} {\bfseries ICHEP2016}
  (2016) 1180} [\href{https://arxiv.org/abs/1611.03007}{{\ttfamily
  1611.03007}}].

\bibitem{Lopez-Val:2014jva}
D.~L\'opez-Val and T.~Robens, \emph{{\ensuremath{\Delta}r and the W-boson mass
  in the singlet extension of the standard model}},
  \href{https://doi.org/10.1103/PhysRevD.90.114018}{\emph{Phys. Rev. D}
  {\bfseries 90} (2014) 114018}
  [\href{https://arxiv.org/abs/1406.1043}{{\ttfamily 1406.1043}}].

\bibitem{CMS:2015hra}
{\scshape CMS} collaboration, \emph{{Search for a Higgs boson in the mass range
  from 145 to 1000 GeV decaying to a pair of W or Z bosons}},
  \href{https://doi.org/10.1007/JHEP10(2015)144}{\emph{JHEP} {\bfseries 10}
  (2015) 144} [\href{https://arxiv.org/abs/1504.00936}{{\ttfamily
  1504.00936}}].

\bibitem{Strassler:2006ri}
M.J.~Strassler and K.M.~Zurek, \emph{{Discovering the Higgs through
  highly-displaced vertices}},
  \href{https://doi.org/10.1016/j.physletb.2008.02.008}{\emph{Phys. Lett. B}
  {\bfseries 661} (2008) 263}
  [\href{https://arxiv.org/abs/hep-ph/0605193}{{\ttfamily hep-ph/0605193}}].

\bibitem{CMS:2016dhk}
{\scshape CMS} collaboration, \emph{{Searches for invisible decays of the Higgs
  boson in pp collisions at $\sqrt{s}$ = 7, 8, and 13 TeV}},
  \href{https://doi.org/10.1007/JHEP02(2017)135}{\emph{JHEP} {\bfseries 02}
  (2017) 135} [\href{https://arxiv.org/abs/1610.09218}{{\ttfamily
  1610.09218}}].

\bibitem{PhysRevLett.44.912}
R.N.~Mohapatra and G.~Senjanovi\ifmmode~\acute{c}\else \'{c}\fi{},
  \emph{Neutrino mass and spontaneous parity nonconservation},
  \href{https://doi.org/10.1103/PhysRevLett.44.912}{\emph{Phys. Rev. Lett.}
  {\bfseries 44} (1980) 912}.

\bibitem{Gell-Mann:1979vob}
M.~Gell-Mann, P.~Ramond and R.~Slansky, \emph{{Complex Spinors and Unified
  Theories}}, {\emph{Conf. Proc. C} {\bfseries 790927} (1979) 315}
  [\href{https://arxiv.org/abs/1306.4669}{{\ttfamily 1306.4669}}].

\bibitem{Wang:2019byi}
W.~Wang and Z.-L.~Han, \emph{{$U(1)_{B-3L_{\alpha}}$ extended scotogenic models
  and single-zero textures of neutrino mass matrices}},
  \href{https://doi.org/10.1103/PhysRevD.101.115040}{\emph{Phys. Rev. D}
  {\bfseries 101} (2020) 115040}
  [\href{https://arxiv.org/abs/1911.00819}{{\ttfamily 1911.00819}}].

\bibitem{Zyla:2020zbs}
{\scshape Particle Data Group} collaboration, \emph{{Review of Particle
  Physics}}, \href{https://doi.org/10.1093/ptep/ptaa104}{\emph{PTEP} {\bfseries
  2020} (2020) 083C01}.

\bibitem{Das:2019fee}
A.~Das, P.S.B.~Dev and N.~Okada, \emph{{Long-lived TeV-scale right-handed
  neutrino production at the LHC in gauged $U(1)_X$ model}},
  \href{https://doi.org/10.1016/j.physletb.2019.135052}{\emph{Phys. Lett. B}
  {\bfseries 799} (2019) 135052}
  [\href{https://arxiv.org/abs/1906.04132}{{\ttfamily 1906.04132}}].

\bibitem{PhysRevD.100.095018}
T.~Hambye, M.H.G.~Tytgat, J.~Vandecasteele and L.~Vanderheyden, \emph{Dark
  matter from dark photons: A taxonomy of dark matter production},
  \href{https://doi.org/10.1103/PhysRevD.100.095018}{\emph{Phys. Rev. D}
  {\bfseries 100} (2019) 095018}.

\bibitem{PhysRevD.102.035028}
R.N.~Mohapatra and N.~Okada, \emph{Dark matter constraints on low mass and
  weakly coupled $b\ensuremath{-}l$ gauge boson},
  \href{https://doi.org/10.1103/PhysRevD.102.035028}{\emph{Phys. Rev. D}
  {\bfseries 102} (2020) 035028}.

\bibitem{Steigman:2012ve}
G.~Steigman, \emph{{Neutrinos And Big Bang Nucleosynthesis}},
  \href{https://doi.org/10.1155/2012/268321}{\emph{Adv. High Energy Phys.}
  {\bfseries 2012} (2012) 268321}
  [\href{https://arxiv.org/abs/1208.0032}{{\ttfamily 1208.0032}}].

\bibitem{Kaplinghat:2013yxa}
M.~Kaplinghat, S.~Tulin and H.-B.~Yu, \emph{{Direct Detection Portals for
  Self-interacting Dark Matter}},
  \href{https://doi.org/10.1103/PhysRevD.89.035009}{\emph{Phys. Rev. D}
  {\bfseries 89} (2014) 035009}
  [\href{https://arxiv.org/abs/1310.7945}{{\ttfamily 1310.7945}}].

\bibitem{PLANCK:2018vyg}
{\scshape Planck} collaboration, \emph{{Planck 2018 results. VI. Cosmological
  parameters}},
  \href{https://doi.org/10.1051/0004-6361/201833910}{\emph{Astron. Astrophys.}
  {\bfseries 641} (2020) A6}
  [\href{https://arxiv.org/abs/1807.06209}{{\ttfamily 1807.06209}}].

\bibitem{Farzan:2017xzy}
Y.~Farzan and M.~Tortola, \emph{{Neutrino oscillations and Non-Standard
  Interactions}}, \href{https://doi.org/10.3389/fphy.2018.00010}{\emph{Front.
  in Phys.} {\bfseries 6} (2018) 10}
  [\href{https://arxiv.org/abs/1710.09360}{{\ttfamily 1710.09360}}].

\bibitem{Proceedings:2019qno}
\emph{{Neutrino Non-Standard Interactions: A Status Report}}, vol.~2, 2019.
\newblock 10.21468/SciPostPhysProc.2.001.

\bibitem{Chang:2016ntp}
J.H.~Chang, R.~Essig and S.D.~McDermott, \emph{{Revisiting Supernova 1987A
  Constraints on Dark Photons}},
  \href{https://doi.org/10.1007/JHEP01(2017)107}{\emph{JHEP} {\bfseries 01}
  (2017) 107} [\href{https://arxiv.org/abs/1611.03864}{{\ttfamily
  1611.03864}}].

\bibitem{Knapen:2017xzo}
S.~Knapen, T.~Lin and K.M.~Zurek, \emph{{Light Dark Matter: Models and
  Constraints}}, \href{https://doi.org/10.1103/PhysRevD.96.115021}{\emph{Phys.
  Rev. D} {\bfseries 96} (2017) 115021}
  [\href{https://arxiv.org/abs/1709.07882}{{\ttfamily 1709.07882}}].

\bibitem{Croon:2020lrf}
D.~Croon, G.~Elor, R.K.~Leane and S.D.~McDermott, \emph{{Supernova Muons: New
  Constraints on $Z$' Bosons, Axions and ALPs}},
  \href{https://doi.org/10.1007/JHEP01(2021)107}{\emph{JHEP} {\bfseries 01}
  (2021) 107} [\href{https://arxiv.org/abs/2006.13942}{{\ttfamily
  2006.13942}}].

\bibitem{Dev:2021qjj}
P.S.B.~Dev, B.~Dutta, K.J.~Kelly, R.N.~Mohapatra and Y.~Zhang, \emph{{Light,
  long-lived B \ensuremath{-} L gauge and Higgs bosons at the DUNE near
  detector}}, \href{https://doi.org/10.1007/JHEP07(2021)166}{\emph{JHEP}
  {\bfseries 07} (2021) 166}
  [\href{https://arxiv.org/abs/2104.07681}{{\ttfamily 2104.07681}}].

\bibitem{Bjorken:2009mm}
J.D.~Bjorken, R.~Essig, P.~Schuster and N.~Toro, \emph{{New Fixed-Target
  Experiments to Search for Dark Gauge Forces}},
  \href{https://doi.org/10.1103/PhysRevD.80.075018}{\emph{Phys. Rev. D}
  {\bfseries 80} (2009) 075018}
  [\href{https://arxiv.org/abs/0906.0580}{{\ttfamily 0906.0580}}].

\bibitem{Blumlein:2011mv}
J.~Blumlein and J.~Brunner, \emph{{New Exclusion Limits for Dark Gauge Forces
  from Beam-Dump Data}},
  \href{https://doi.org/10.1016/j.physletb.2011.05.046}{\emph{Phys. Lett. B}
  {\bfseries 701} (2011) 155}
  [\href{https://arxiv.org/abs/1104.2747}{{\ttfamily 1104.2747}}].

\bibitem{Blumlein:2013cua}
J.~Bl\"umlein and J.~Brunner, \emph{{New Exclusion Limits on Dark Gauge Forces
  from Proton Bremsstrahlung in Beam-Dump Data}},
  \href{https://doi.org/10.1016/j.physletb.2014.02.029}{\emph{Phys. Lett. B}
  {\bfseries 731} (2014) 320}
  [\href{https://arxiv.org/abs/1311.3870}{{\ttfamily 1311.3870}}].

\bibitem{Essig:2010gu}
R.~Essig, R.~Harnik, J.~Kaplan and N.~Toro, \emph{{Discovering New Light States
  at Neutrino Experiments}},
  \href{https://doi.org/10.1103/PhysRevD.82.113008}{\emph{Phys. Rev. D}
  {\bfseries 82} (2010) 113008}
  [\href{https://arxiv.org/abs/1008.0636}{{\ttfamily 1008.0636}}].

\bibitem{SHiP:2015vad}
{\scshape SHiP} collaboration, \emph{{A facility to Search for Hidden Particles
  (SHiP) at the CERN SPS}},  \href{https://arxiv.org/abs/1504.04956}{{\ttfamily
  1504.04956}}.

\bibitem{Feng:2017vli}
J.L.~Feng, I.~Galon, F.~Kling and S.~Trojanowski, \emph{{Dark Higgs bosons at
  the ForwArd Search ExpeRiment}},
  \href{https://doi.org/10.1103/PhysRevD.97.055034}{\emph{Phys. Rev. D}
  {\bfseries 97} (2018) 055034}
  [\href{https://arxiv.org/abs/1710.09387}{{\ttfamily 1710.09387}}].

\bibitem{FASER:2018eoc}
{\scshape FASER} collaboration, \emph{{FASER\textquoteright{}s physics reach
  for long-lived particles}},
  \href{https://doi.org/10.1103/PhysRevD.99.095011}{\emph{Phys. Rev. D}
  {\bfseries 99} (2019) 095011}
  [\href{https://arxiv.org/abs/1811.12522}{{\ttfamily 1811.12522}}].

\bibitem{FASER:2018bac}
{\scshape FASER} collaboration, \emph{{Technical Proposal for FASER: ForwArd
  Search ExpeRiment at the LHC}},
  \href{https://arxiv.org/abs/1812.09139}{{\ttfamily 1812.09139}}.

\bibitem{FASER:2019aik}
{\scshape FASER} collaboration, \emph{{FASER: ForwArd Search ExpeRiment at the
  LHC}},  \href{https://arxiv.org/abs/1901.04468}{{\ttfamily 1901.04468}}.

\bibitem{Belle-II:2018jsg}
{\scshape Belle-II} collaboration, \emph{{The Belle II Physics Book}},
  \href{https://doi.org/10.1093/ptep/ptz106}{\emph{PTEP} {\bfseries 2019}
  (2019) 123C01} [\href{https://arxiv.org/abs/1808.10567}{{\ttfamily
  1808.10567}}].

\bibitem{BaBar:2014zli}
{\scshape BaBar} collaboration, \emph{{Search for a Dark Photon in $e^+e^-$
  Collisions at BaBar}},
  \href{https://doi.org/10.1103/PhysRevLett.113.201801}{\emph{Phys. Rev. Lett.}
  {\bfseries 113} (2014) 201801}
  [\href{https://arxiv.org/abs/1406.2980}{{\ttfamily 1406.2980}}].

\bibitem{BaBar:2017tiz}
{\scshape BaBar} collaboration, \emph{{Search for Invisible Decays of a Dark
  Photon Produced in ${e}^{+}{e}^{-}$ Collisions at BaBar}},
  \href{https://doi.org/10.1103/PhysRevLett.119.131804}{\emph{Phys. Rev. Lett.}
  {\bfseries 119} (2017) 131804}
  [\href{https://arxiv.org/abs/1702.03327}{{\ttfamily 1702.03327}}].

\bibitem{COHERENT:2020iec}
{\scshape COHERENT} collaboration, \emph{{First Measurement of Coherent Elastic
  Neutrino-Nucleus Scattering on Argon}},
  \href{https://doi.org/10.1103/PhysRevLett.126.012002}{\emph{Phys. Rev. Lett.}
  {\bfseries 126} (2021) 012002}
  [\href{https://arxiv.org/abs/2003.10630}{{\ttfamily 2003.10630}}].

\bibitem{Coloma:2015pih}
P.~Coloma, B.A.~Dobrescu, C.~Frugiuele and R.~Harnik, \emph{{Dark matter beams
  at LBNF}}, \href{https://doi.org/10.1007/JHEP04(2016)047}{\emph{JHEP}
  {\bfseries 04} (2016) 047}
  [\href{https://arxiv.org/abs/1512.03852}{{\ttfamily 1512.03852}}].

\bibitem{Coloma:2020gfv}
P.~Coloma, M.C.~Gonzalez-Garcia and M.~Maltoni, \emph{{Neutrino oscillation
  constraints on U(1)' models: from non-standard interactions to long-range
  forces}}, \href{https://doi.org/10.1007/JHEP01(2021)114}{\emph{JHEP}
  {\bfseries 01} (2021) 114}
  [\href{https://arxiv.org/abs/2009.14220}{{\ttfamily 2009.14220}}].

\bibitem{Amaral:2020tga}
D.W.P.d.~Amaral, D.G.~Cerdeno, P.~Foldenauer and E.~Reid, \emph{{Solar neutrino
  probes of the muon anomalous magnetic moment in the gauged $
  \mathrm{U}{(1)}_{L_{\mu }-{L}_{\tau }} $}},
  \href{https://doi.org/10.1007/JHEP12(2020)155}{\emph{JHEP} {\bfseries 12}
  (2020) 155} [\href{https://arxiv.org/abs/2006.11225}{{\ttfamily
  2006.11225}}].

\bibitem{Bellini:2011rx}
G.~Bellini et~al., \emph{{Precision measurement of the 7Be solar neutrino
  interaction rate in Borexino}},
  \href{https://doi.org/10.1103/PhysRevLett.107.141302}{\emph{Phys. Rev. Lett.}
  {\bfseries 107} (2011) 141302}
  [\href{https://arxiv.org/abs/1104.1816}{{\ttfamily 1104.1816}}].

\bibitem{Borexino:2017rsf}
{\scshape Borexino} collaboration, \emph{{First Simultaneous Precision
  Spectroscopy of $pp$, $^7$Be, and $pep$ Solar Neutrinos with Borexino
  Phase-II}}, \href{https://doi.org/10.1103/PhysRevD.100.082004}{\emph{Phys.
  Rev. D} {\bfseries 100} (2019) 082004}
  [\href{https://arxiv.org/abs/1707.09279}{{\ttfamily 1707.09279}}].

\bibitem{Heeck:2018nzc}
J.~Heeck, M.~Lindner, W.~Rodejohann and S.~Vogl, \emph{{Non-Standard Neutrino
  Interactions and Neutral Gauge Bosons}},
  \href{https://doi.org/10.21468/SciPostPhys.6.3.038}{\emph{SciPost Phys.}
  {\bfseries 6} (2019) 038} [\href{https://arxiv.org/abs/1812.04067}{{\ttfamily
  1812.04067}}].

\bibitem{Kamada:2015era}
A.~Kamada and H.-B.~Yu, \emph{{Coherent Propagation of PeV Neutrinos and the
  Dip in the Neutrino Spectrum at IceCube}},
  \href{https://doi.org/10.1103/PhysRevD.92.113004}{\emph{Phys. Rev. D}
  {\bfseries 92} (2015) 113004}
  [\href{https://arxiv.org/abs/1504.00711}{{\ttfamily 1504.00711}}].

\bibitem{Escudero:2019gzq}
M.~Escudero, D.~Hooper, G.~Krnjaic and M.~Pierre, \emph{{Cosmology with A Very
  Light L$_{\mu}$ \ensuremath{-} L$_{\tau}$ Gauge Boson}},
  \href{https://doi.org/10.1007/JHEP03(2019)071}{\emph{JHEP} {\bfseries 03}
  (2019) 071} [\href{https://arxiv.org/abs/1901.02010}{{\ttfamily
  1901.02010}}].

\bibitem{PhysRevLett.66.3117}
S.R.~Mishra, S.A.~Rabinowitz, C.~Arroyo, K.T.~Bachmann, R.E.~Blair, C.~Foudas
  et~al., \emph{Neutrino tridents and w-z interference},
  \href{https://doi.org/10.1103/PhysRevLett.66.3117}{\emph{Phys. Rev. Lett.}
  {\bfseries 66} (1991) 3117}.

\bibitem{Altmannshofer:2019zhy}
W.~Altmannshofer, S.~Gori, J.~Mart\'\i{}n-Albo, A.~Sousa and M.~Wallbank,
  \emph{{Neutrino Tridents at DUNE}},
  \href{https://doi.org/10.1103/PhysRevD.100.115029}{\emph{Phys. Rev. D}
  {\bfseries 100} (2019) 115029}
  [\href{https://arxiv.org/abs/1902.06765}{{\ttfamily 1902.06765}}].

\bibitem{Davier:2017zfy}
M.~Davier, A.~Hoecker, B.~Malaescu and Z.~Zhang, \emph{{Reevaluation of the
  hadronic vacuum polarisation contributions to the Standard Model predictions
  of the muon $g-2$ and ${\alpha (m_Z^2)}$ using newest hadronic cross-section
  data}}, \href{https://doi.org/10.1140/epjc/s10052-017-5161-6}{\emph{Eur.
  Phys. J. C} {\bfseries 77} (2017) 827}
  [\href{https://arxiv.org/abs/1706.09436}{{\ttfamily 1706.09436}}].

\bibitem{Davier:2019can}
M.~Davier, A.~Hoecker, B.~Malaescu and Z.~Zhang, \emph{{A new evaluation of the
  hadronic vacuum polarisation contributions to the muon anomalous magnetic
  moment and to $\mathbf{\boldsymbol\alpha(m_Z^2)}$}},
  \href{https://doi.org/10.1140/epjc/s10052-020-7792-2}{\emph{Eur. Phys. J. C}
  {\bfseries 80} (2020) 241}
  [\href{https://arxiv.org/abs/1908.00921}{{\ttfamily 1908.00921}}].

\bibitem{Muong-2:2006rrc}
{\scshape Muon g-2} collaboration, \emph{{Final Report of the Muon E821
  Anomalous Magnetic Moment Measurement at BNL}},
  \href{https://doi.org/10.1103/PhysRevD.73.072003}{\emph{Phys. Rev. D}
  {\bfseries 73} (2006) 072003}
  [\href{https://arxiv.org/abs/hep-ex/0602035}{{\ttfamily hep-ex/0602035}}].

\bibitem{Aoyama:2020ynm}
T.~Aoyama et~al., \emph{{The anomalous magnetic moment of the muon in the
  Standard Model}},
  \href{https://doi.org/10.1016/j.physrep.2020.07.006}{\emph{Phys. Rept.}
  {\bfseries 887} (2020) 1} [\href{https://arxiv.org/abs/2006.04822}{{\ttfamily
  2006.04822}}].

\bibitem{Muong-2:2021ojo}
{\scshape Muon g-2} collaboration, \emph{{Measurement of the Positive Muon
  Anomalous Magnetic Moment to 0.46 ppm}},
  \href{https://doi.org/10.1103/PhysRevLett.126.141801}{\emph{Phys. Rev. Lett.}
  {\bfseries 126} (2021) 141801}
  [\href{https://arxiv.org/abs/2104.03281}{{\ttfamily 2104.03281}}].

\bibitem{DONuT:2007bsg}
{\scshape DONuT} collaboration, \emph{{Final tau-neutrino results from the
  DONuT experiment}},
  \href{https://doi.org/10.1103/PhysRevD.78.052002}{\emph{Phys. Rev. D}
  {\bfseries 78} (2008) 052002}
  [\href{https://arxiv.org/abs/0711.0728}{{\ttfamily 0711.0728}}].

\bibitem{Ghoshal:2019pab}
A.~Ghoshal, A.~Giarnetti and D.~Meloni, \emph{{On the role of the $\nu_{\tau}$
  appearance in DUNE in constraining standard neutrino physics and beyond}},
  \href{https://doi.org/10.1007/JHEP12(2019)126}{\emph{JHEP} {\bfseries 12}
  (2019) 126} [\href{https://arxiv.org/abs/1906.06212}{{\ttfamily
  1906.06212}}].

\bibitem{Ghoshal:2020hyo}
A.~Ghoshal, A.~Giarnetti and D.~Meloni, \emph{{Neutrino Invisible Decay at
  DUNE: a multi-channel analysis}},
  \href{https://doi.org/10.1088/1361-6471/abdfab}{\emph{J. Phys. G} {\bfseries
  48} (2021) 055004} [\href{https://arxiv.org/abs/2003.09012}{{\ttfamily
  2003.09012}}].

\bibitem{Ma:1998dp}
E.~Ma and D.P.~Roy, \emph{{Phenomenology of the $B$ - 3L($\tau$) gauge boson}},
  \href{https://doi.org/10.1103/PhysRevD.58.095005}{\emph{Phys. Rev. D}
  {\bfseries 58} (1998) 095005}
  [\href{https://arxiv.org/abs/hep-ph/9806210}{{\ttfamily hep-ph/9806210}}].

\bibitem{SHiP:2020sos}
{\scshape SHiP} collaboration, \emph{{SND@LHC}},
  \href{https://arxiv.org/abs/2002.08722}{{\ttfamily 2002.08722}}.

\bibitem{Kling:2020iar}
F.~Kling, \emph{{Probing light gauge bosons in tau neutrino experiments}},
  \href{https://doi.org/10.1103/PhysRevD.102.015007}{\emph{Phys. Rev. D}
  {\bfseries 102} (2020) 015007}
  [\href{https://arxiv.org/abs/2005.03594}{{\ttfamily 2005.03594}}].

\bibitem{Aoki:2017spj}
S.~Aoki et~al., \emph{{Study of tau-neutrino production at the CERN SPS}},
  \href{https://arxiv.org/abs/1708.08700}{{\ttfamily 1708.08700}}.

\bibitem{DsTau:2019wjb}
{\scshape DsTau} collaboration, \emph{{DsTau: Study of tau neutrino production
  with 400 GeV protons from the CERN-SPS}},
  \href{https://doi.org/10.1007/JHEP01(2020)033}{\emph{JHEP} {\bfseries 01}
  (2020) 033} [\href{https://arxiv.org/abs/1906.03487}{{\ttfamily
  1906.03487}}].

\bibitem{DUNE:2015lol}
{\scshape DUNE} collaboration, \emph{{Long-Baseline Neutrino Facility (LBNF)
  and Deep Underground Neutrino Experiment (DUNE)}: {Conceptual Design Report,
  Volume 2: The Physics Program for DUNE at LBNF}},
  \href{https://arxiv.org/abs/1512.06148}{{\ttfamily 1512.06148}}.

\bibitem{DUNE:2016hlj}
{\scshape DUNE} collaboration, \emph{{Long-Baseline Neutrino Facility (LBNF)
  and Deep Underground Neutrino Experiment (DUNE)}: {Conceptual Design Report,
  Volume 1: The LBNF and DUNE Projects}},
  \href{https://arxiv.org/abs/1601.05471}{{\ttfamily 1601.05471}}.

\bibitem{Berryman:2019dme}
J.M.~Berryman, A.~de~Gouvea, P.J.~Fox, B.J.~Kayser, K.J.~Kelly and J.L.~Raaf,
  \emph{{Searches for Decays of New Particles in the DUNE Multi-Purpose Near
  Detector}}, \href{https://doi.org/10.1007/JHEP02(2020)174}{\emph{JHEP}
  {\bfseries 02} (2020) 174}
  [\href{https://arxiv.org/abs/1912.07622}{{\ttfamily 1912.07622}}].

\bibitem{Bakhti:2018avv}
P.~Bakhti, Y.~Farzan and M.~Rajaee, \emph{{Secret interactions of neutrinos
  with light gauge boson at the DUNE near detector}},
  \href{https://doi.org/10.1103/PhysRevD.99.055019}{\emph{Phys. Rev. D}
  {\bfseries 99} (2019) 055019}
  [\href{https://arxiv.org/abs/1810.04441}{{\ttfamily 1810.04441}}].

\bibitem{PhysRevLett.116.042501}
A.J.~Krasznahorkay, M.~Csatl\'os, L.~Csige, Z.~G\'acsi, J.~Guly\'as, M.~Hunyadi
  et~al., \emph{Observation of anomalous internal pair creation in
  $^{8}\mathrm{Be}$: A possible indication of a light, neutral boson},
  \href{https://doi.org/10.1103/PhysRevLett.116.042501}{\emph{Phys. Rev. Lett.}
  {\bfseries 116} (2016) 042501}.

\bibitem{Feng:2016jff}
J.L.~Feng, B.~Fornal, I.~Galon, S.~Gardner, J.~Smolinsky, T.M.P.~Tait et~al.,
  \emph{{Protophobic Fifth-Force Interpretation of the Observed Anomaly in
  $^8$Be Nuclear Transitions}},
  \href{https://doi.org/10.1103/PhysRevLett.117.071803}{\emph{Phys. Rev. Lett.}
  {\bfseries 117} (2016) 071803}
  [\href{https://arxiv.org/abs/1604.07411}{{\ttfamily 1604.07411}}].

\bibitem{Nardi:2018cxi}
E.~Nardi, C.D.R.~Carvajal, A.~Ghoshal, D.~Meloni and M.~Raggi, \emph{{Resonant
  production of dark photons in positron beam dump experiments}},
  \href{https://doi.org/10.1103/PhysRevD.97.095004}{\emph{Phys. Rev. D}
  {\bfseries 97} (2018) 095004}
  [\href{https://arxiv.org/abs/1802.04756}{{\ttfamily 1802.04756}}].

\bibitem{LHCb:2015nkv}
{\scshape LHCb} collaboration, \emph{{Search for hidden-sector bosons in $B^0
  \!\to K^{*0}\mu^+\mu^-$ decays}},
  \href{https://doi.org/10.1103/PhysRevLett.115.161802}{\emph{Phys. Rev. Lett.}
  {\bfseries 115} (2015) 161802}
  [\href{https://arxiv.org/abs/1508.04094}{{\ttfamily 1508.04094}}].

\bibitem{LHCb:2016awg}
{\scshape LHCb} collaboration, \emph{{Search for long-lived scalar particles in
  $B^+ \to K^+ \chi (\mu^+\mu^-)$ decays}},
  \href{https://doi.org/10.1103/PhysRevD.95.071101}{\emph{Phys. Rev. D}
  {\bfseries 95} (2017) 071101}
  [\href{https://arxiv.org/abs/1612.07818}{{\ttfamily 1612.07818}}].

\bibitem{AristizabalSierra:2015vqb}
D.~Aristizabal~Sierra, F.~Staub and A.~Vicente, \emph{{Shedding light on the
  $b\to s$ anomalies with a dark sector}},
  \href{https://doi.org/10.1103/PhysRevD.92.015001}{\emph{Phys. Rev. D}
  {\bfseries 92} (2015) 015001}
  [\href{https://arxiv.org/abs/1503.06077}{{\ttfamily 1503.06077}}].

\bibitem{Datta:2017pfz}
A.~Datta, J.~Liao and D.~Marfatia, \emph{{A light $Z^\prime$ for the $R_K$
  puzzle and nonstandard neutrino interactions}},
  \href{https://doi.org/10.1016/j.physletb.2017.02.058}{\emph{Phys. Lett. B}
  {\bfseries 768} (2017) 265}
  [\href{https://arxiv.org/abs/1702.01099}{{\ttfamily 1702.01099}}].

\bibitem{Datta:2017ezo}
A.~Datta, J.~Kumar, J.~Liao and D.~Marfatia, \emph{{New light mediators for the
  $R_K$ and $R_{K^*}$ puzzles}},
  \href{https://doi.org/10.1103/PhysRevD.97.115038}{\emph{Phys. Rev. D}
  {\bfseries 97} (2018) 115038}
  [\href{https://arxiv.org/abs/1705.08423}{{\ttfamily 1705.08423}}].

\bibitem{Sala:2017ihs}
F.~Sala and D.M.~Straub, \emph{{A New Light Particle in B Decays?}},
  \href{https://doi.org/10.1016/j.physletb.2017.09.072}{\emph{Phys. Lett. B}
  {\bfseries 774} (2017) 205}
  [\href{https://arxiv.org/abs/1704.06188}{{\ttfamily 1704.06188}}].

\bibitem{Kang:2019vng}
Z.~Kang and Y.~Shigekami, \emph{{$(g-2)_{\mu}$ versus flavor changing neutral
  current induced by the light $(B-L)_{\mu\tau}$ boson}},
  \href{https://doi.org/10.1007/JHEP11(2019)049}{\emph{JHEP} {\bfseries 11}
  (2019) 049} [\href{https://arxiv.org/abs/1905.11018}{{\ttfamily
  1905.11018}}].

\bibitem{Borah:2020swo}
D.~Borah, L.~Mukherjee and S.~Nandi, \emph{{Low scale U(1)$_{X}$ gauge symmetry
  as an origin of dark matter, neutrino mass and flavour anomalies}},
  \href{https://doi.org/10.1007/JHEP12(2020)052}{\emph{JHEP} {\bfseries 12}
  (2020) 052} [\href{https://arxiv.org/abs/2007.13778}{{\ttfamily
  2007.13778}}].

\bibitem{Cen:2021iwv}
J.-Y.~Cen, Y.~Cheng, X.-G.~He and J.~Sun, \emph{{Flavor Specific
  $U(1)_{B_q-L_\mu}$ Gauge Model for Muon $g-2$ and $b \to s \bar \mu \mu$
  Anomalies}},  \href{https://arxiv.org/abs/2104.05006}{{\ttfamily
  2104.05006}}.

\end{thebibliography}\endgroup

\end{document}